\documentclass[a4paper,fleqn,usenatbib,onecolumn]{mnras}
\usepackage[T1]{fontenc}
\usepackage{ae,aecompl}
\usepackage{graphicx}	
\usepackage{amsmath}	
\usepackage{amssymb}	
\usepackage{booktabs}
\usepackage{hyperref}
\usepackage{bm}

\newcommand{\kms}{\text{km~s$^{-1}$}}
\newcommand{\gpercmsq}{\text{g~cm$^{-2}$}}
\newcommand{\gpercmcu}{\text{g~cm$^{-3}$}}
\newcommand{\msun}{\text{M$_\odot$}}
\newcommand{\msunperyr}{\text{M$_\odot$~yr$^{-1}$}}

\defcitealias{2016ApJ...825...14S}{SF16}
\defcitealias{2017arXiv170104627Z}{ZS17}

\title[Ring and gap formation]{Rings and gaps produced by variable magnetic disk winds and avalanche accretion streams: I. Axisymmetric resistive MHD simulations}

\author[Suriano et al.]{
Scott S. Suriano$^{1}$\thanks{E-mail: suriano@virginia.edu},
Zhi-Yun Li$^{1}$, Ruben Krasnopolsky$^{2}$, and Hsien Shang$^{2}$
\\
$^{1}$Department of Astronomy, University of Virginia, Charlottesville, VA 22904, USA \\
$^{2}$Academia Sinica, Institute of Astronomy and Astrophysics, Taipei 10617, Taiwan\\
}

\date{Accepted XXX. Received YYY; in original form ZZZ}

\pubyear{2017}

\begin{document}
\label{firstpage}
\pagerange{\pageref{firstpage}--\pageref{lastpage}}
\maketitle

\begin{abstract}
Rings and gaps are being observed in an increasing number of disks around young stellar objects. We illustrate the formation of such radial structures through idealized, 2D (axisymmetric) resistive MHD simulations of coupled disk-wind systems threaded by a relatively weak poloidal magnetic field (plasma-$\beta \sim 10^3$). We find two distinct modes of accretion depending on the resistivity and field strength. A small resistivity or high field strength promotes the development of rapidly infalling `avalanche accretion streams' in a vertically extended disk envelope that dominates the dynamics of the system, especially the mass accretion. The streams are suppressed in simulations with larger resistivities or lower field strengths, where most of the accretion instead occurs through a laminar disk. In these simulations, the disk accretion is driven mainly by a slow wind that is typically accelerated by the pressure gradient from a predominantly toroidal magnetic field. Both wind-dominated and stream-dominated modes of accretion create prominent features in the surface density distribution of the disk, including rings and gaps, with a strong spatial variation of the magnetic flux relative to the mass. Regions with low mass-to-flux ratios accrete quickly, leading to the development of gaps, whereas regions with higher mass-to-flux ratios tend to accrete more slowly, allowing matter to accumulate and form dense rings. In some cases, avalanche accretion streams are observed to produce dense rings directly through continuous feeding. We discuss the implications of ring and gap formation driven by winds and streams on grain growth and planet formation.
\end{abstract}



\section{Introduction}

Disks around young stellar objects (YSOs) are the birthplaces of
planets. Planet formation and evolution thus depend critically on the
properties of these disks. Given the large number of super-Earths and 
mini-Neptunes discovered by \textit{Kepler} at distances between $\sim 0.1$ 
and $\sim 1$~au from their host stars \citep{2015ARA&A..53..409W}, there 
is strong motivation for studying the structure and evolution of the inner ($\lesssim1$~au)
parts of protostellar disks. This will be the focus of 
our investigation. 

The inner circumstellar 
disk plays an important role in launching outflows. Jets and 
winds are ubiquitously observed in YSOs and have a long history of
observational and theoretical study (see \citealt{2014prpl.conf..451F} for a review and references therein). 
It is thought that such outflows are driven by rapidly rotating magnetic field lines, although
where the outflow-driving field lines are anchored remains unclear.   
One school of thought posits that the field lines are trapped at the 
inner edge of the disk, giving rise to so-called `X-winds' 
\citep{2000prpl.conf..789S}. Another suggests that they are distributed over a more  
extended region of the inner disk, driving a `disk wind' \citep{2000prpl.conf..759K}.
Part of the 
outflow may also be attributed to an accretion-enhanced stellar wind (e.g., \citealt{2008ApJ...678.1109M}). Some evidence favoring disk winds comes from the observed gradients in the 
line-of-sight velocity measured across optical jets (e.g., \citealt{2000ApJ...537L..49B}) and molecular outflows (e.g., \citealt{2009A&A...494..147L}). If such gradients arise from rotation in a 
magnetocentrifugal wind \citep{1982MNRAS.199..883B}, the implied 
outflow launching region would extend to au scales or larger 
(e.g., \citealt{2003ApJ...590L.107A,2007prpl.conf..231R,2016Natur.540..406B}; however, see Lee et al., \textit{submitted}, for SiO observations of HH 212 that provide evidence for X-winds). 
The outflow would carry angular momentum away from the disk in 
a launching region that is directly relevant to the formation of the 
terrestrial planets in the Solar system and a considerable fraction of the 
exoplanets discovered by \textit{Kepler}.

The idea of magnetic wind-driven disk evolution has been discussed 
in the literature for a long time (e.g., \citealt{1983ApJ...274..677P,1993ApJ...410..218W,2008A&A...479..481C,2010MNRAS.401..479K,2016ApJ...821...80B,2016A&A...596A..74S}, and references therein). One way to quantify the effects that 
winds have on disks is to construct global, coupled wind-disk 
solutions to the MHD equations. Many of the early investigations along this line adopted 
simplifying self-similarity assumptions (e.g., \citealt{1993ApJ...410..218W,1995ApJ...444..848L,1997A&A...319..340F,2011MNRAS.412.1162S}). 
These semi-analytic treatments have been important for illuminating the 
basic mechanics of the coupled system, including how a small fraction 
of the accreting material is peeled off of the disk surface and 
ejected along the rapidly rotating field lines as a wind, how the 
wind removes angular momentum from the disk and drives it to accrete, 
and how the accretion and ejection processes depend on the non-ideal 
MHD effects (Ohmic dissipation, ambipolar diffusion, 
and the Hall effect) that are expected to occur in the lightly ionized YSO
disks. Self-similar 
solutions require
that all physical quantities, such as the density, magnetic field 
strength, and the coefficients for non-ideal MHD effects, must 
scale with the radius as power-laws with specific indices. This 
requirement limits the applicability of the self-similar solutions 
to realistic systems. We seek to relax the self-similarity ansatz 
through a series of MHD simulations that incorporate non-ideal effects.

Simulations of coupled wind-disk systems have been carried out 
by several groups (e.g., \citealt{2002ApJ...565.1035K,2002ApJ...581..988C,2004ApJ...601...90C,2007A&A...469..811Z,2010A&A...512A..82M,2012MNRAS.420.2020L,2012ApJ...757...65S,2013ApJ...768....5C,2014ApJ...793...31S,2016ApJ...825...14S}). These simulations confirmed and 
extended the earlier semi-analytic results, finding self-consistent 
wind-disk solutions for a wide range of disk masses and magnetic 
field distributions. However, the focus of such simulations has 
typically been on the launching of outflows from the disk and on their 
propagation to large distances. 
We are thus motivated to 
start a long-term program to investigate the wind-disk system 
that will eventually include all three non-ideal MHD effects 
using the ZeusTW MHD code \citep{2011ApJ...733...54K,2011ApJ...738..180L}, focusing in particular on the structure of the disk in such a system. 
As a first step, we start with the simplest case of a resistive wind-disk system 
under the assumptions of axisymmetry (2D) and reflection symmetry across the disk midplane.

Resistive disk winds under such an idealized geometry have been 
studied by several groups, especially \citet{2016ApJ...825...14S} (herein,~\citetalias{2016ApJ...825...14S}), 
who were able to determine wind properties (such as its mass 
loss rate, energy, and angular momentum) for disks with a wide 
range of magnetic field strengths. The emphasis of such
work has been on quasi-steady state wind-disk solutions, which 
enable the computation of several well-known conserved quantities 
along each magnetic flux surface (e.g., \citealt{2010LNP...794..233S}) 
that can be checked against the semi-analytic solutions. 
However, a steady state can only be obtained when outward diffusion 
of the magnetic field in the disk is balanced by inward advection of the field,
which occurs only under rather restrictive 
conditions for the prescribed magnetic diffusivity. Under more 
general conditions, the magnetic winds remain highly variable, 
driving non-steady disk accretion even 
for 2D (axisymmetric) systems; they are likely to be exacerbated 
in 3D simulations. Such non-steady state phenomena are the focus
of this investigation. We find that the disk structure can be completely 
reshaped from its initial state by even a 
weak initial magnetic field, often with rapidly accreting streams 
developing near the disk surface and rings and gaps 
developing near the disk midplane. 

The accretion streams are a form of the magnetorotational instability (MRI) channel flows \citep{1991ApJ...376..214B,1992ApJ...400..595H,1994ApJ...432..213G}. They are the result of the runaway magnetic braking of an infalling stream, which is linked to more slowly rotating material at larger radii. We term them `avalanche accretion streams', motivated by the work of \citet{1998ApJ...508..186K} who find similar structures in ideal MHD simulations of thick AGN disks (see also \citealt{1994ApJ...433..746S,1996ApJ...461..115M,2002PASJ...54..121K,2009ApJ...707..428B}). We find that the accretion streams become more numerous and more important dynamically as the disk material becomes better coupled to the magnetic field. Their rapid formation and disruption forms a thick envelope of chaotic infall and outflow motions above the disk, which is intimately related to the so-called `coronal accretion' found recently by \citet{2017arXiv170104627Z} (herein,~\citetalias{2017arXiv170104627Z}) in global (3D) ideal MHD simulations of thin disks with net vertical magnetic flux (see also \citealt{2009ApJ...707..428B}). These investigations highlight the possible importance of vigorous accretion in a vertically extended structure outside the dense disk on disk evolution.

The formation of rings and gaps 
on sub-au scales strongly influences how 
dust grains are trapped and thus affects how the grains grow into  
planetesimals and ultimately planets. Although our simulations are limited 
to the inner ($\sim 0.01-1$ au scale) disk, the same mechanism of ring and 
gap formation through variable disk winds and avalanche accretion streams 
should operate at larger radii as well, where rings and gaps have now been 
observed in a number of disks 
\citep{2015ApJ...808L...3A,2016ApJ...820L..40A,2016ApJ...818L..16Z,2016PhRvL.117y1101I}. 
The 2D simulations presented in this study serve as an 
illustration of these generic mechanisms, although they are likely to generate rings and gaps more easily than 3D simulations, because of the assumed axisymmetry. Nevertheless, they provide a 
starting point for future explorations that will include more 
detailed disk microphysics and less restrictive geometry.

The rest of the paper is organized as follows. In Section~\ref{sec:setup}, we describe the simulation setup. In Section~\ref{sec:ref}, we present the results of a reference simulation. In Section~\ref{sec:param}, we analyze how the simulation outcome depends on three key dimensionless parameters that control the disk magnetic field strength, resistivity and disk thickness. In Section~\ref{sec:dis}, we discuss this work in the context of previous studies and examine its implications. The results are summarized in Section~\ref{sec:conc}.

\section{Problem setup}\label{sec:setup}

\subsection{MHD equations}
We use the ZeusTW code \citep{2010ApJ...716.1541K} to solve the
time-dependent, resistive magnetohydrodynamic (MHD) equations in axisymmetric spherical
coordinates ($r,\theta,\phi$). The equations solved are as follows:
\begin{equation}
\frac{\partial \rho}{\partial t} + \nabla \cdot \left( \rho \bm{v} \right) = 0, 
\end{equation}

\begin{equation}
\rho\frac{\partial\bm{v}}{\partial t} + \rho\left(\bm{v}\cdot\nabla\right)\bm{v} = -\nabla P + \bm{J}\times\bm{B}/c - \rho\nabla\Phi_g,
\end{equation}

\begin{equation}
\frac{\partial \bm{B}}{\partial t} =  \nabla \times \left(\bm{v} \times \bm{B} - \eta \nabla \times \bm{B}\right),
\end{equation}

\begin{equation}
\frac{\partial e}{\partial t} + \nabla \cdot \left(e \bm{v} \right) = -P \nabla \cdot \bm{v}.
\end{equation}
The current density is $\bm{J}=(c/4\pi)\nabla\times\bm{B}$, the
internal energy is $e=P/(\gamma-1)$, and $\eta$ is the resistivity.  The remaining parameters have
their usual definitions. When referring to cylindrical coordinates, 
we will use the notation ($R,\phi,z$) such that $R=r\sin{\theta}$ 
and $z=r\cos{\theta}$.
 
\subsection{Initial conditions}

The simulation domain is separated into two regions: a thin, cold, rotating disk orbiting a 1~\msun~central source at the grid origin and a stationary, hot corona above the disk. We
choose $\gamma=1.01$ so that the disk and corona regions are locally isothermal.

\subsubsection{Disk}

The geometrically thin disk is characterized by the
dimensionless parameter $\epsilon = h/r = c_s/v_K \ll 1$, where $h$ 
is the disk scale height, $c_s$ is the sound speed, and
$v_K$ is the Keplerian speed. The disk is limited to the equatorial region
where the polar angle $\theta \in {[\pi/2 - \theta_0,\pi/2]}$, with 
disk (half) opening angle set to $\theta_0=\arctan(2\epsilon)$. The 
disk density takes the form of a radial power law multiplied by a Gaussian 
function of $z/r=\cos\theta$,
\begin{equation}
\rho_d(r,\theta) = \rho_{0} \left(\frac{r}{r_0}\right)^{-\alpha} \exp \left(-\frac{\cos^2\theta}{2 \epsilon^2}\right),
\end{equation}
as dictated by hydrostatic balance. The subscript `0' refers to values on the disk midplane at the 
inner radial boundary. For all simulations shown this paper, we use $\alpha=3/2$. The choice of power-law exponent is consistent with sub-millimeter observations of Class II sources that find surface density power-law exponents of $0.4-1.1$ \citep{2010ApJ...723.1241A}. The disk pressure is set as
\begin{equation}
P_d(r,\theta)=\rho_d c_s^2,
\end{equation}
with $c_s = \epsilon v_K$. The radial pressure gradient causes the equilibrium rotational velocity $v_\phi$ to be slightly sub-Keplerian as
\begin{equation}
v_\phi =  v_K \sqrt{1-(1+\alpha)\epsilon^2}.
\end{equation}

\subsubsection{Corona}
We require that the hydrostatic corona is initially in pressure balance with the disk surface. This constraint sets the density drop from the disk surface to the corona as $(1+\alpha)\epsilon^2$. Therefore, the coronal density and pressure are
\begin{equation}
\rho_{c}(r)=\rho_{0} \epsilon^2(1+\alpha)\exp\left[-\frac{\cos^2\theta_0}{2 \epsilon^2}\right] \left(\frac{r}{r_0}\right)^{-\alpha}=\rho_{c,0}\left(\frac{r}{r_0}\right)^{-\alpha},
\end{equation}
\begin{equation}
P_c(r)= \rho_c v_K^2/(1+\alpha).
\end{equation}
The coronal sound speed is then $c_{s,c}=v_K/\sqrt{1+\alpha}$.

\subsubsection{Magnetic field}
To ensure that the magnetic field is divergence-free initially,
we set the magnetic field components using the magnetic flux function
$\Psi$ as in \citet{2007A&A...469..811Z},
\begin{equation}
\Psi(r,\theta) = \frac{4}{3}r_0^2 B_{p,0}\left(\frac{r\sin\theta}{r_0}\right)^{3/4} \frac{m^{5/4}}{\left(m^{2}+\cot^2\theta\right)^{5/8}}\label{eq:psi},
\end{equation}
where the $B_{p,0}$ is the poloidal field strength on the midplane at $r_0$, and the parameter $m$ controls the bending of the field. Since varying $m$ from 0.1 to 1 has little effect on the long-term disk or wind magnetic field structure \citep{2014ApJ...793...31S}, we use $m=0.5$ for all simulations presented in this work. The initial magnetic field components are then calculated as
\begin{equation}
B_r=\frac{1}{r^2\sin{\theta}}\frac{\partial\Psi}{\partial\theta},
\end{equation}
\begin{equation}
B_\theta=-\frac{1}{r\sin\theta}\frac{\partial\Psi}{\partial r}.
\end{equation}

\subsubsection{Resistivity}
For illustrative purposes in this initial study, we adopt a spatially and temporally constant resistivity 
$\eta(\bm{r},t)=\eta$; more refined treatments, including one based on detailed ionization calculation, are 
postponed to future investigations. This implementation differs from the resistivity profile used in other disk wind simulations where $\eta\propto hv_A$ ($v_A$ is the Alfv\'en speed) 
inside the disk and $\eta=0$ outside (e.g., \citealt{2014ApJ...793...31S}).
To quantify the relative importance of the prescribed resistivity, we define a dimensionless magnetic
diffusivity parameter $D \equiv\eta/(hc_s)$ (where $h$ is the disk scale height and $c_s$ is the disk sound speed) as in \citet{1995ApJ...444..848L}, 
similar to the well-known $\alpha$-parameter of \citet{1973A&A....24..337S}. This diffusivity parameter is
initially proportional to $r^{-1/2}$ with $D_0=0.16$ at the inner edge
of the disk in the reference
simulation, corresponding to an ionization fraction of $x_e=2.3\times10^{-12}$ at 1 au \citep{1994ApJ...421..163B}. This is not far from the minimum ionization fraction due to the radioactive decay of $^{26}$Al alone ($x_e \approx 1.9\times 10^{-13}$ according to \citealt{2011ARA&A..49..195A}). The resistivity adopted in our reference simulation is large enough to suppress the fastest growing MRI mode near the disk midplane \citep{1996ApJ...457..798J}. However, the MRI may still operate, especially near the disk surface where the plasma-$\beta$ is lower and the critical resistivity needed for MRI suppression (which is inversely proportional to $\beta$) is higher.

\subsection{Grid}\label{sec:grid}
The equations are solved for $r\in{[0.02,10]}$~au and
$\theta\in{[0,\pi/2]}$, where $\theta=0$ corresponds to the polar axis and $\theta=\pi/2$ to the disk midplane.
We choose $r_0=0.02~\rm{au}\sim 4~\rm{R}_\odot$
as the inner radius because it is approximately the inner radius of
a protostellar disk truncated by the magnetosphere of 
a classical T Tauri star \citep{2016ARA&A..54..135H}. It yields an inner orbital period
$t_0\approx 1$~d. For this initial study, we focus on only 
one hemisphere and assume that the system has reflection symmetry
across the disk midplane. This assumption will be relaxed in future 
investigations. In the radial direction, a
`ratioed' grid is used, such that $dr_{i+1}/dr_i$ is constant and
$r_{i+1}=r_i+dr_i$. The grid spacing at the inner edge is set such that
$dr_0=2.3r_0d\theta$. The grid is uniform in the $\theta$-coordinate
for most simulations presented in the work, with a 
resolution of $n_r\times n_\theta = 400\times360$. This results in 24 grid cells from the disk midplane to surface (at two disk
scale heights) in the reference run. The fastest growing (ideal MHD) MRI wavelength would be marginally resolved at the disk midplane with $\lambda=3.3rd\theta$ initially \citep{1991ApJ...376..214B}.

In the simulation called x2-grid, the $\theta$-grid is separated into
two regions: a uniform grid in the disk region ($\theta\in{[\pi/3,\pi/2]}$),
and a ratioed grid in the corona ($\theta\in{[0,\pi/3]}$), with $d\theta$ matched at the boundary
and increasing towards the polar axis. The uniform region contains 240
cells, thereby decreasing $d\theta$ by a factor of two compared to the
reference run. The non-uniform grid section has 120 cells, so the total number of cells in the $\theta$-direction remains 360. We decrease the value of $dr_0$ in order to have the ratio of $\frac{dr_0}{r_0d\theta}$ be the same in all runs. The value of
$d\theta$ increases to $1.3^\circ$ at the polar axis for this ratioed grid. In this case, the fastest growing (ideal MHD) MRI wavelength would be resolved by about 7 cells.

\subsection{Boundary conditions}\label{sec:bc}

The inner radial boundary is separated into two regions. For $\theta
\in{[0,\pi/2-\arctan(4\epsilon)]}$, mass is injected into the
simulation domain with $\rho=\rho_{c,0}$ and $v=10~\text{km}~\text{s}^{-1}$ \citep{1999ApJ...526..631K}. The
injection boundary is used to lower the Alf\'ven speed in polar region
close to the inner boundary and thereby increase the simulation time step, 
as the region would quickly be evacuated by gravitational infall if the standard outflow
boundary was used. The total mass that enters the 
simulation domain through the inner boundary is small compared to both 
the initial mass in the simulation and the mass that is eventually carried
through the outer boundary via the disk wind. 
It thus has little influence on the dynamics of
the simulation. The remaining section of the inner radial boundary,
$\theta>\pi/2-\arctan(4\epsilon)$, and the outer radial boundary both use
the standard outflow condition, as usual in Zeus codes.

The reflection boundary condition is used on the polar axis ($\theta=0$)
and the equatorial plane ($\theta=\pi/2$). The $\phi$-component of the 
magnetic field is set to zero on the polar axis. The assumed 
reflection symmetry across the midplane ($\theta=\pi/2$) demands that 
$B_\phi=0$ at this boundary. We also set $B_\phi$ to vanish on the inner 
radial boundary because it is not rotating. 

\section{Reference model}\label{sec:ref}

\begin{figure}
\centering
\includegraphics[width=1.0\textwidth]{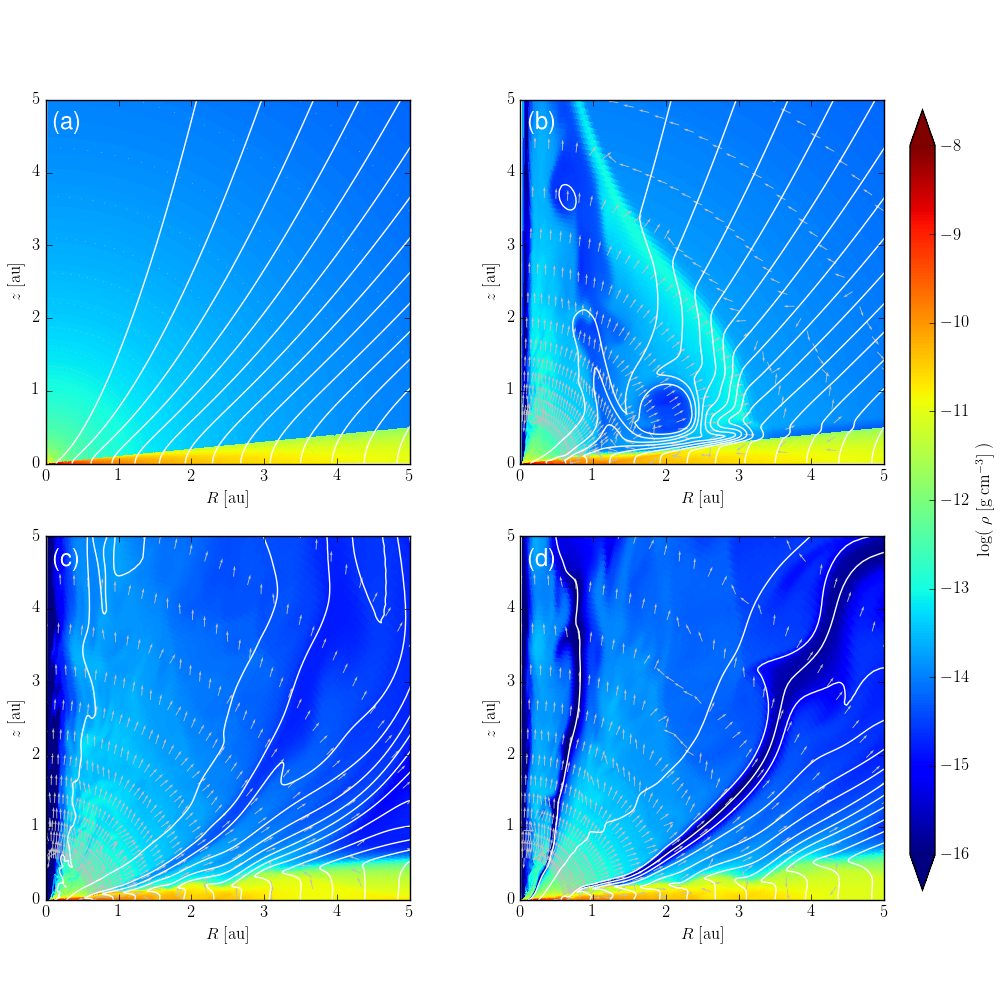}
\caption{A representative (`reference') axisymmetric simulation of a coupled disk-wind system. Shown is the mass volume density (logarithmically spaced colour contours in units of \gpercmcu), the poloidal magnetic field lines (white), and the poloidal velocity unit vectors (grey). Panels (a)-(d) corresponding to simulation times of 0, 150, 1200, and 1800 inner orbital periods, respectively. (See the supplementary material in the online journal for an animated version of this figure.)}
\label{fig:snapshot}
\end{figure}

We start by discussing the result of a `reference' simulation. 
It is used to highlight the 
salient features of the coupled wind-disk system, and serves as a
benchmark against which other simulations with different parameters 
will be compared in Section~\ref{sec:param}. Of the parameters that will be 
changed in Section~\ref{sec:param}, the reference model has $\epsilon=0.05$, an initial magnetic 
field strength characterized by plasma-$\beta$ of $10^3$ on the 
disk midplane (corresponding to 0.27~G at 1 au), and a dimensionless magnetic diffusivity parameter of $D_0=0.16$.
This simulation runs for 1820 inner orbital periods ($\sim 5$~yr). Fig.~\ref{fig:snapshot}
shows the initial conditions for the system and snapshots at three representative times.

\subsection{Global evolution}

The initially weak poloidal magnetic field is
wound up by differential rotation between the (nearly) Keplerian
disk and the static corona, which inflates a bubble of strong toroidal
magnetic field that expands outward against the coronal material. This
winding operates fastest near the inner edge of the disk, where the 
orbital period is the shortest. As a result, the outflow propagates
with the highest speed near the polar axis. By $t=150 t_0$ (Fig.~\ref{fig:snapshot}(b)), 
the outflow
has reached the outer edge of the computation domain along the axis,
but remains confined by the initial coronal material away from the
polar region. By $t=1200 t_0$ (Fig.~\ref{fig:snapshot}(c)), most of the initial coronal material has been completely swept out of the computational domain. Beyond this time, the effects of the initial corona on the
coupled wind-disk system should be relatively small, and for the inner 
part of the system (where much of our analysis is focused; see Section~\ref{sec:zoneI} below), 
the initial corona ceases to be important at an even earlier 
time. The wind-disk system shows large spatial and temporal variability 
throughout the simulation. This variability is reflected in Fig.~\ref{fig:snapshot}(d) at $t=1800 t_0$; 
its appearance is broadly similar to that in panel (c), but there are important differences
such as the appearance of low-density wind `channels' that are more prominent in the outflow 
of the former than the latter. The variability of the outflow is intimately 
tied to the structures that develop in the disk.

\subsection{Outflow}

\begin{figure}
\centering
\includegraphics[width=1.0\textwidth]{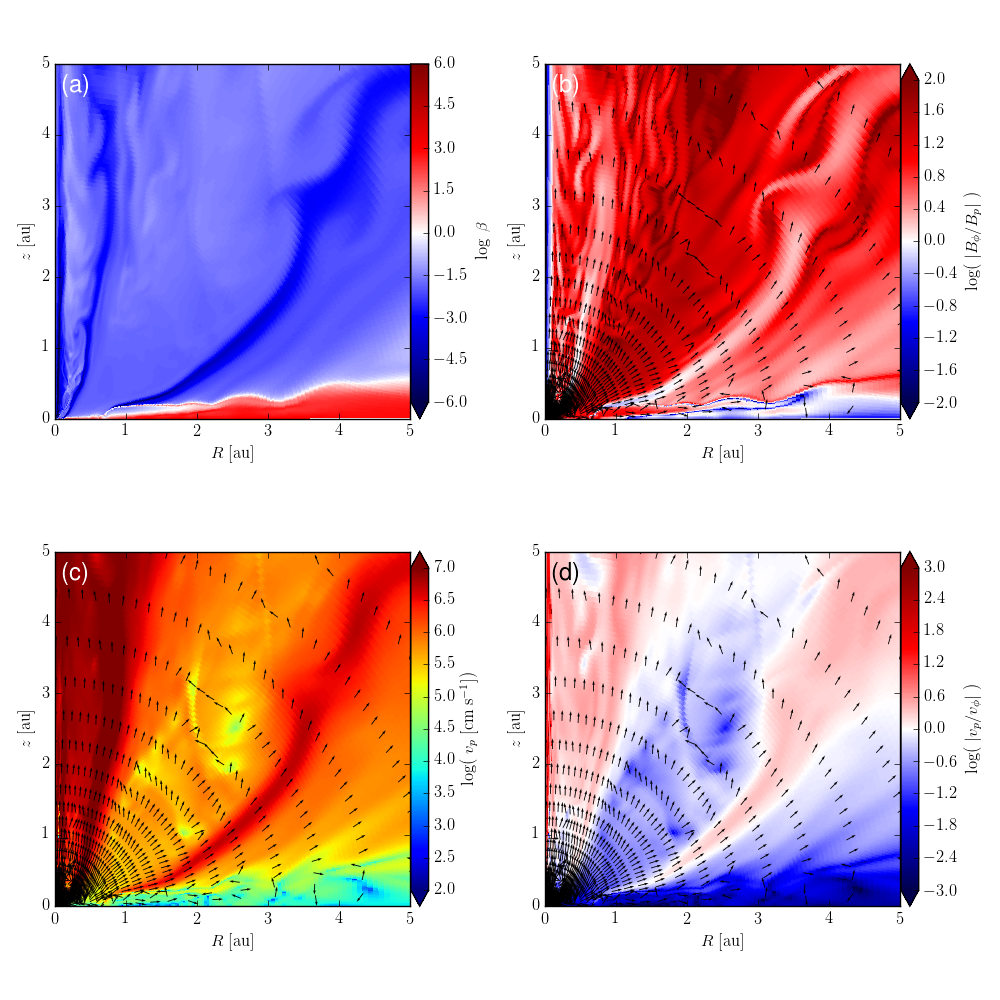}
\caption{The reference simulation at time $t=1800~t_0$. The logarithmically spaced colour contours show (a) the plasma-$\beta$, (b) the ratio of the toroidal to the poloidal magnetic field components, $\vert B_\phi/B_p\vert$, (c) the poloidal velocity (cm~s$^{-1}$), and (d) the ratio of poloidal to the toroidal velocity components, $\vert v_p/v_\phi\vert$.
Panels (b)-(d) show poloidal velocity unit vectors (black).}
\label{fig:beta}
\end{figure}

Even though the initial magnetic field at the disk midplane is
rather weak ($\beta=10^3$), it is still 
able to drive a powerful outflow. 
Unlike in the disk, the pressure in the 
outflow is dominated by the magnetic field rather than the thermal 
gas, as illustrated in Fig.~\ref{fig:beta}(a). This 
dominant magnetic pressure is provided mostly by the toroidal
component $B_\phi$, as shown in panel (b) where the ratio of the toroidal and poloidal components
$\vert B_\phi/B_p\vert$ is plotted. This ratio is 
spatially inhomogeneous, with a filamentary appearance. 
The dominance of the toroidal field component suggests that the outflow is driven
mostly by the magnetic pressure gradient, as is expected for a
relatively weak initial magnetic field (e.g., \citetalias{2016ApJ...825...14S}) or a heavy mass loading (e.g., \citealt{2005ApJ...630..945A}). 
The exceptions are a narrow region near the polar axis and 
two filaments at roughly 
$\theta=10^\circ$ and $80^\circ$. The magnetic field lines that run through the polar axis region are connected
to the inner radial boundary, which is
assumed to be non-rotating; the lack of
a toroidal field component here is therefore expected. The other two 
filaments are physically meaningful and correspond to the two low-density `channels' that are 
prominent in the density map of Fig.~\ref{fig:snapshot}(d). In these channels, the 
thermal pressure is completely dominated by the magnetic pressure,
with a plasma $\beta$ less than $10^{-3}$ (see Fig.~\ref{fig:beta}(a)). In other
words, their field lines are much more lightly 
mass-loaded than in the rest of the outflow. As a result, the magnetic field is able to accelerate the 
mass to a much higher speed than in the denser surrounding 
regions (see Fig.~\ref{fig:beta}(c)). 

The distribution of the poloidal velocity is plotted in panel (c) of 
Fig.~\ref{fig:beta}. The outflow in the outer low-density channel reaches a 
speed of $\sim 50$~\kms, whereas that in the inner channel
is much faster, reaching up to 200~\kms. The outflow close to the 
axis can also reach a relatively high speed of $\sim 50-100$~\kms. 
However, over most of the simulation volume, especially to the right of the 
inner channel, the poloidal outflow speed is rather low, 
typically a factor of two or more smaller than the the rotation speed 
and well below the local free-fall speed (see Fig.~\ref{fig:beta}(d)). 
Except for the channels and polar axis region, the magnetic and velocity 
fields of the slowly-expanding, low plasma-$\beta$ outflow are dominated by their toroidal components.

\begin{figure}
\centering
\includegraphics[width=1.0\textwidth]{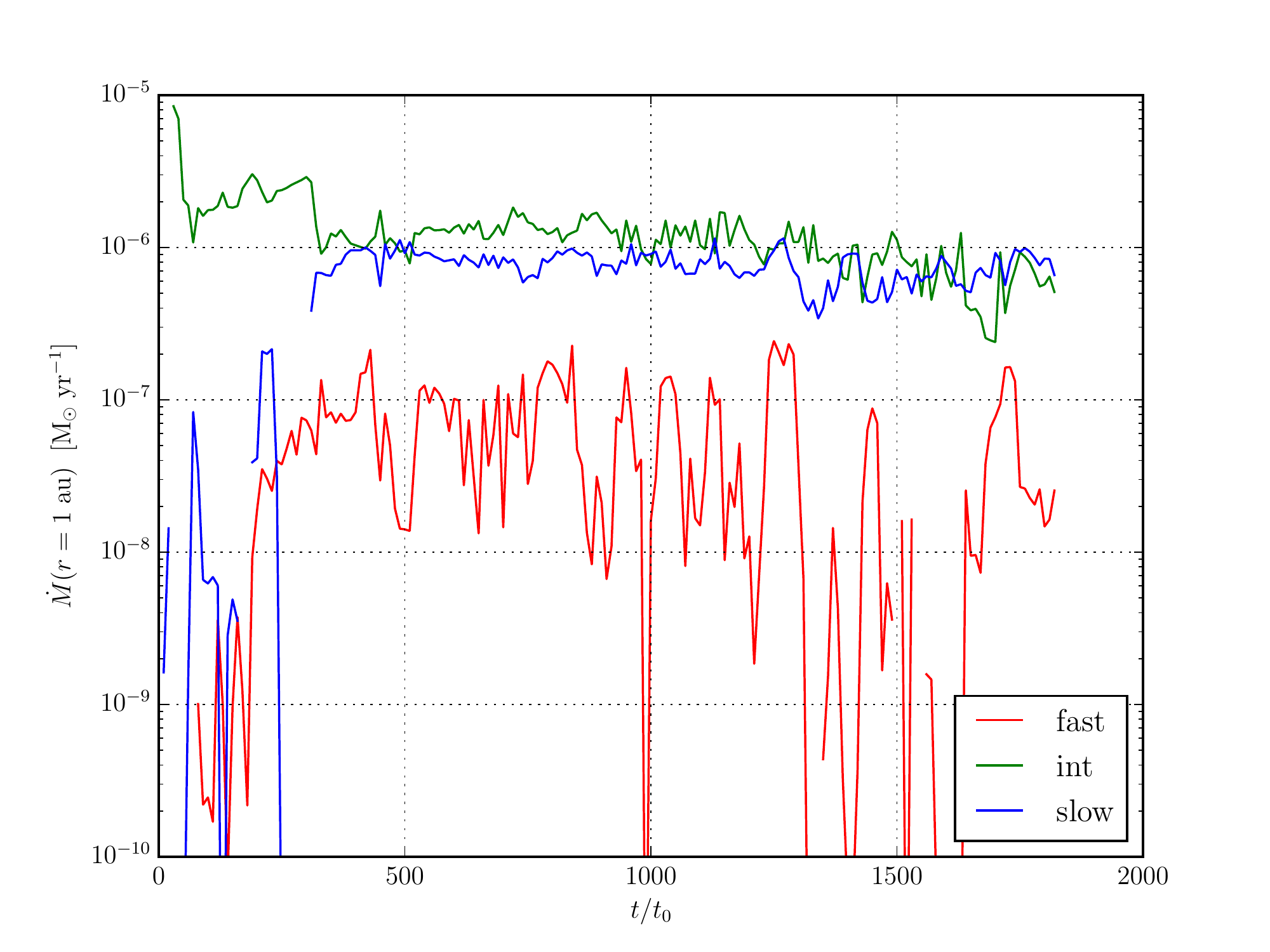}
\caption{Mass outflow rate (\msunperyr) through hemisphere of $r=1$~au as a function of time in the reference simulation. The mass loss rate is separated into three velocity components. The fast velocity component ($v_r>100$~\kms) is shown in red, the intermediate velocity component (10~\kms~$<v_r<100$~\kms) in green, and the slow velocity component (1~\kms~$<v_r<10$~\kms) in blue.}
\label{fig:mdot_var}
\end{figure}

Even though the outflow is slow in general, there is disk material that gets accelerated to a high speed. In
Fig.~\ref{fig:mdot_var}, we plot the mass flux of the outflowing material
with a radial velocity greater than 100~\kms~as a function of time through
a hemisphere of radius of $r=1$~au. It is clear that
the fastest component of the outflow is highly variable 
in time. The mass outflow rate routinely changes by an order of magnitude on timescales of $t\lesssim100t_0$, 
and can sometimes drop by four orders of magnitude on similar timescales.
This variability may be essential for generating internal
shocks that are required to keep the outflow heated to a relatively 
high temperature and visible through, e.g, optical forbidden lines 
(e.g., \citealt{2002ApJ...564..853S}). The average mass loss 
rate of this fast component is approximately $5\times10^{-8}$~\msunperyr, 
which is at the high end of the mass loss rate inferred in 
the classical T Tauri jets (e.g., \citealt{2014prpl.conf..451F}).
This fast, massive outflow component is remarkable in 
light of the fact that the disk is only weakly magnetized, at least 
initially with a midplane plasma-$\beta$ of $10^3$.

The fast component is only a relatively minor component of the outflow produced in the 
reference simulation. This is illustrated in Fig.~\ref{fig:mdot_var},
where the mass loss rate is plotted as a function of time for the slow
(1~\kms $< v_r <$ 10~\kms) and intermediate-speed (10~\kms$ < v_r< $
100~\kms) components of the outflow. Both components have mass loss rates of order
$10^{-6}$~\msunperyr, which is at least an order of
magnitude higher than the mass flux of the fast component.
Such a massive outflow is expected to strongly affect -- indeed, control -- the 
disk dynamics, as we demonstrate in Section~\ref{sec:zoneII}.

\subsection{Disk-wind connection}

\begin{figure}
	\includegraphics[width=0.95\columnwidth]{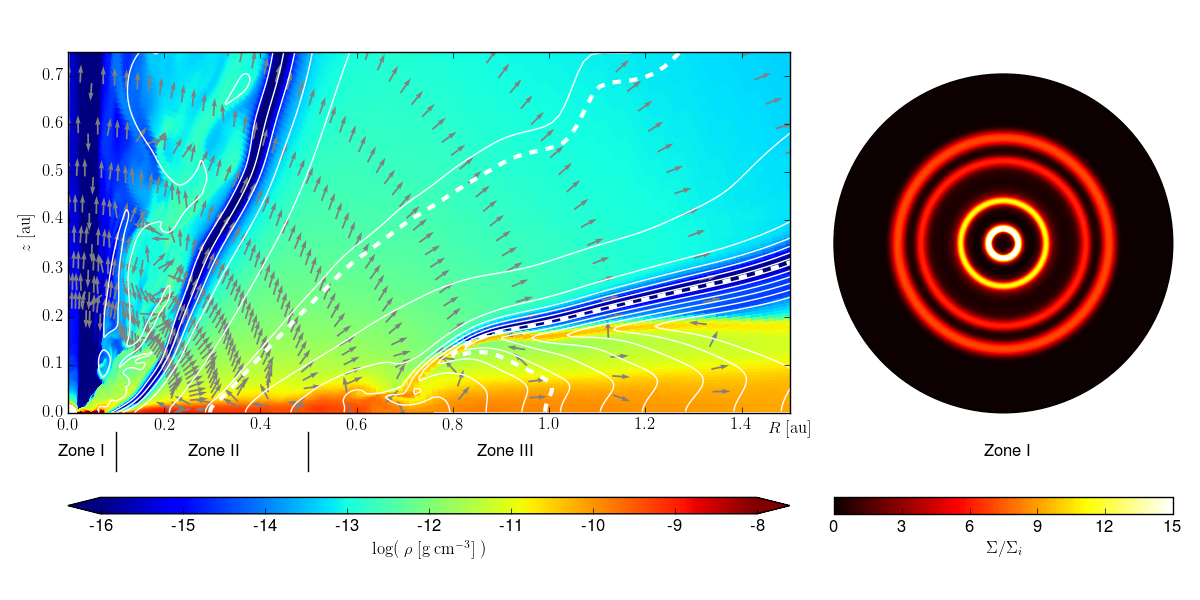}
    \caption{The reference simulation at $t=1800 t_0$. Left: The mass volume density (\gpercmcu) is shown in logarithmically spaced colour contours. Magnetic field lines are shown in white and the two dashed lines show the field lines with midplane footpoints of 0.3 and 1.0~au (along which the quantities in Fig.~\ref{fig:path_0.3au} and \ref{fig:path_1au} are plotted). The grey arrows denote the velocity field and show that the bulk of the disk material is expanding under the surface accretion stream beyond a radius of $\sim 0.7$~au. Right: The `face-on' axisymmetric surface density normalized to its initial distribution is shown in colour contours for Zone I ($r\leq0.1$~au).}
    \label{fig:den_4panel}
\end{figure}

Fig.~\ref{fig:den_4panel} shows the density distribution and magnetic field lines for the final time frame of the reference simulation, as in panel (d) of Fig.~\ref{fig:snapshot} and \ref{fig:beta}, but now zoomed in on the inner part of disk where the disk structure is most strongly affected by the outflow. As expected, the modification of disk structure starts near the inner edge, where the rotation period is the shortest. By $t=1800~t_0$ (Fig.~\ref{fig:den_4panel}), three distinct regions have been established: (1) an inner region ($r<0.1$~au) with a highly variable mass distribution that contains prominent dense rings and gaps, (2) an intermediate radius region (0.1~au~$<r<0.5$~au) with a much smoother mass density distribution, and (3) an outer region ($r>0.5$~au) where the disk `puffs up' due to the presence of a dense surface layer -- the `avalanche accretion stream.' We label these regions Zone I, II, and III, respectively, and discuss each in detail below. We save the discussion of the most variable region, Zone I, until Section~\ref{sec:zoneI}.

\subsubsection{Zone II: Steady wind-driven accretion}\label{sec:zoneII}

\begin{figure}
    \begin{minipage}{0.33\linewidth}
	\includegraphics[width=1.0\columnwidth]{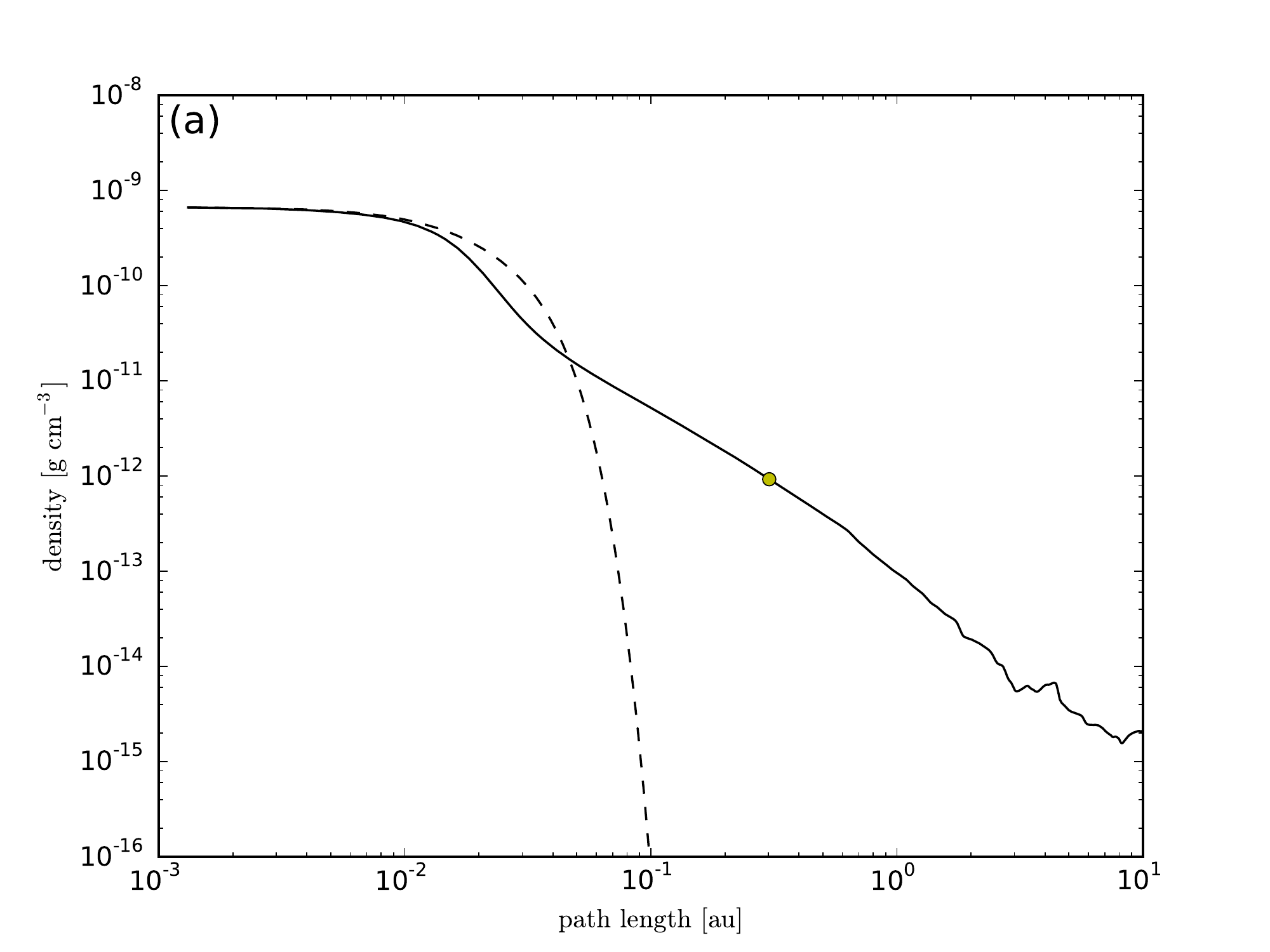}
	\end{minipage}
	\begin{minipage}{0.33\linewidth}
	\includegraphics[width=1.0\columnwidth]{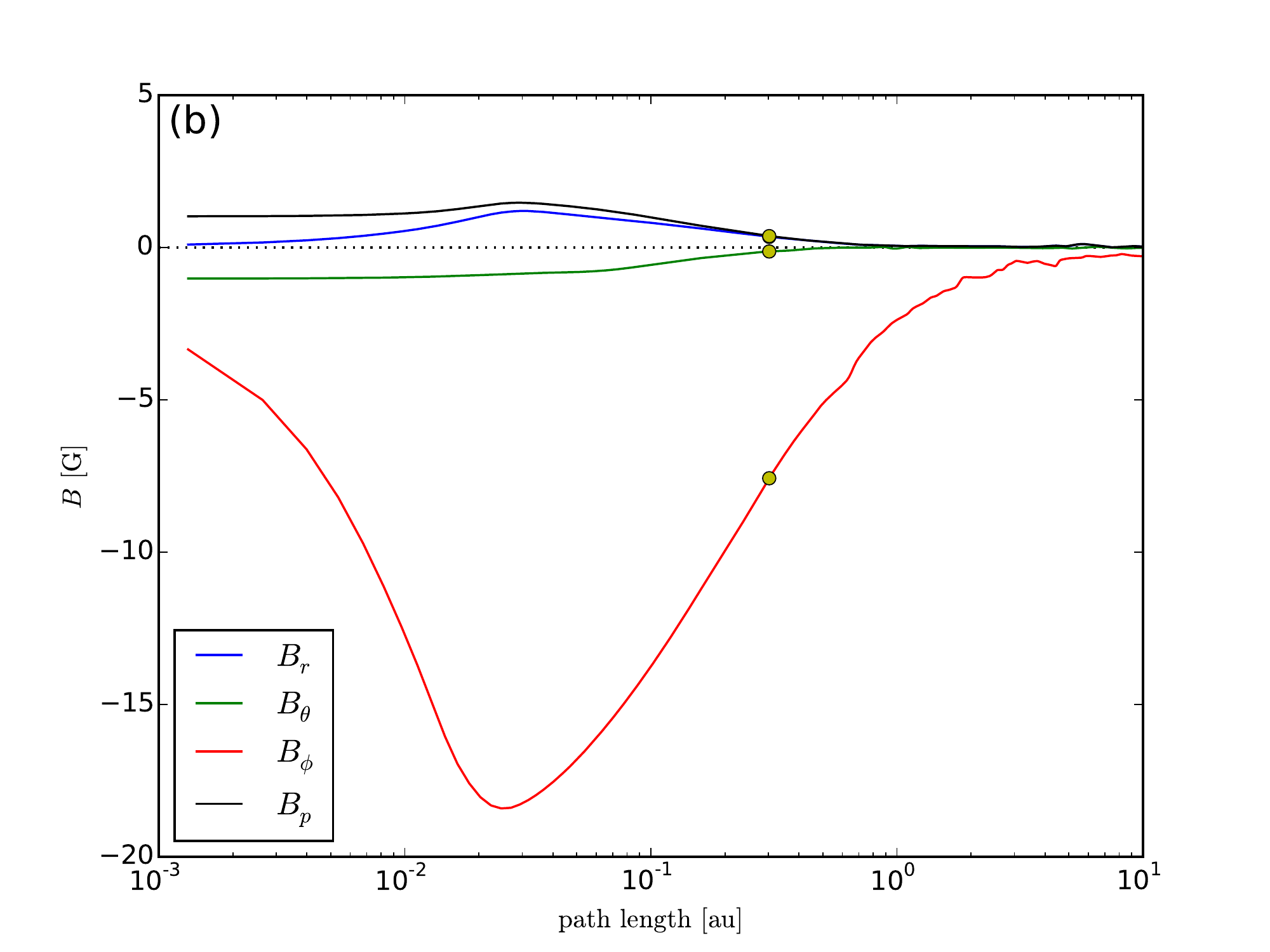}
	\end{minipage}
	\begin{minipage}{0.33\linewidth}
	\includegraphics[width=1.0\columnwidth]{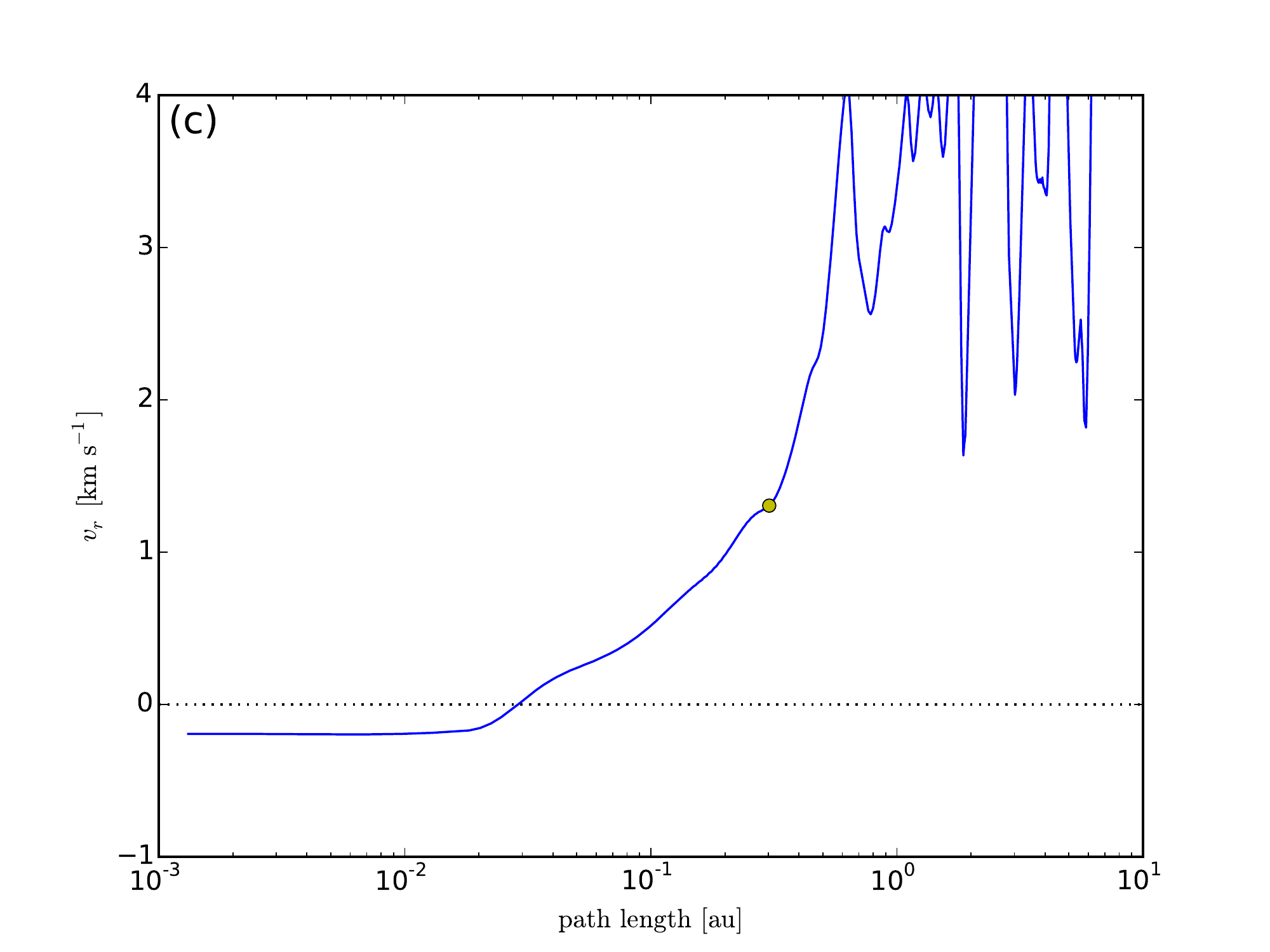}
	\end{minipage}
	\begin{minipage}{0.33\linewidth}
	\includegraphics[width=1.0\columnwidth]{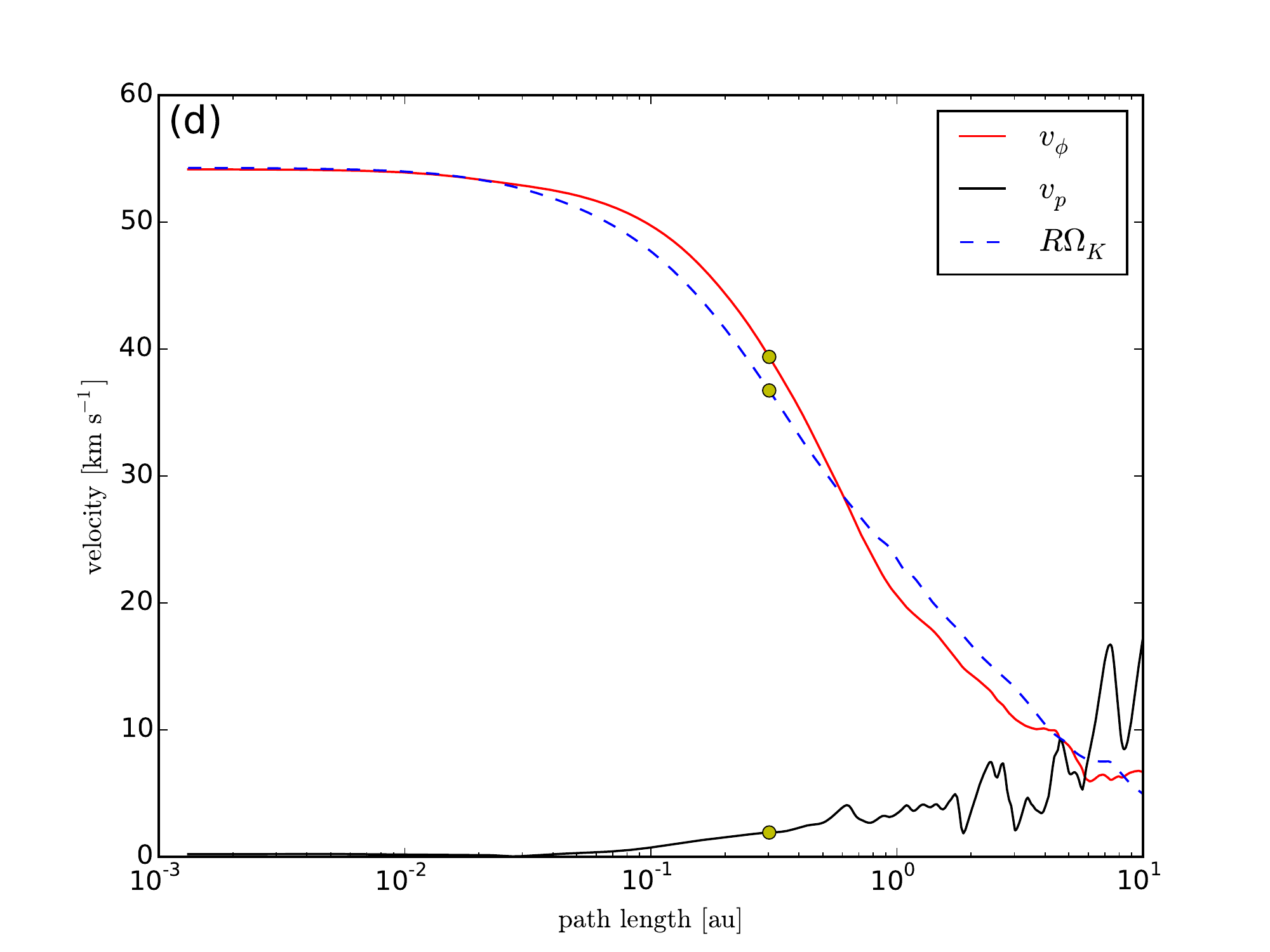}
	\end{minipage}
	\begin{minipage}{0.33\linewidth}
	\includegraphics[width=1.0\columnwidth]{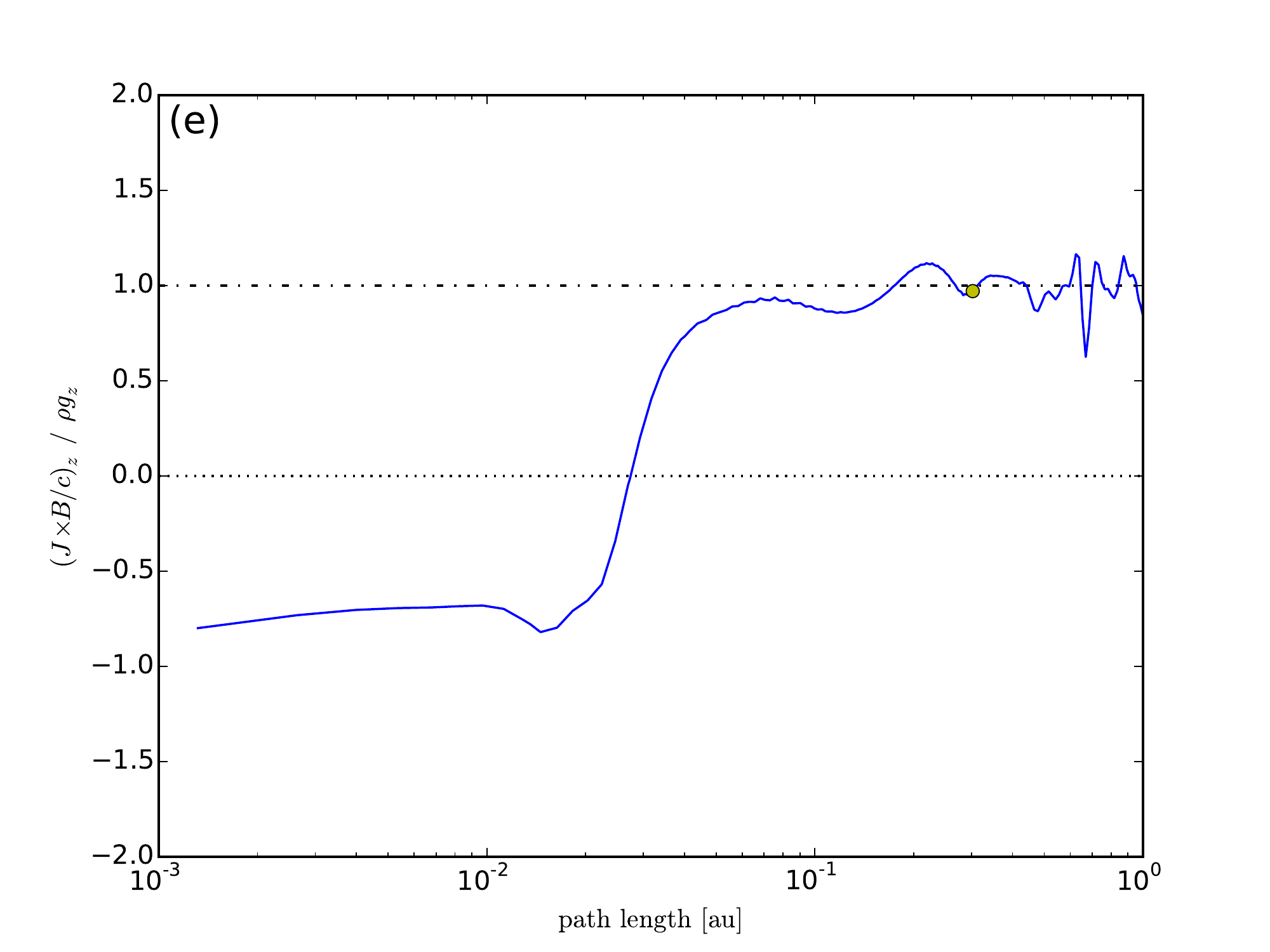}
	\end{minipage}
	\begin{minipage}{0.33\linewidth}
	\includegraphics[width=1.0\columnwidth]{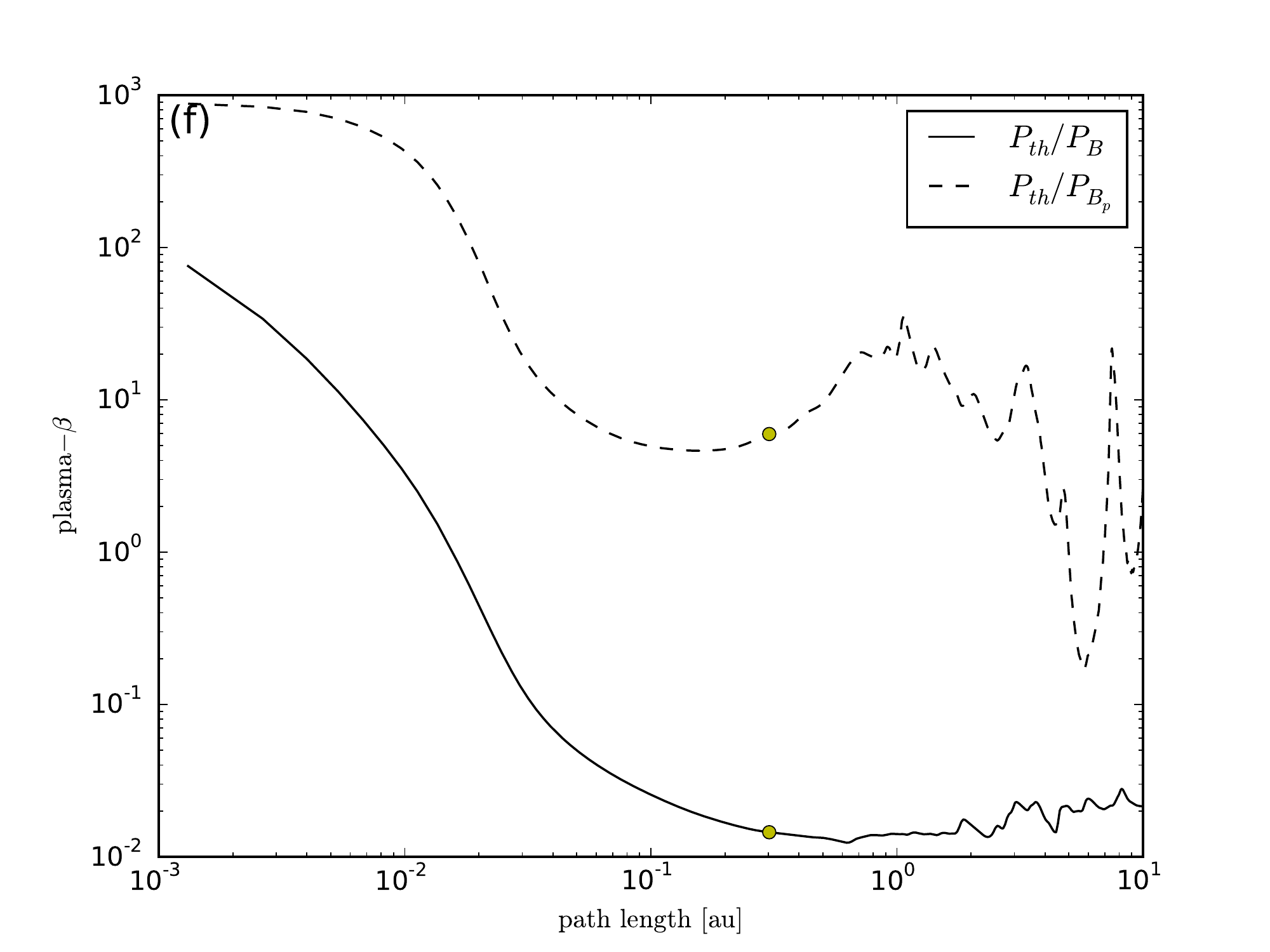}
	\end{minipage}
    \caption{Physical quantities plotted along a poloidal magnetic field line as a function of the distance along the line. The representative field line has a footpoint at $r=0.3$~au (Zone II) and can be seen in Fig.~\ref{fig:den_4panel} (white dashed line). Yellow circles show the sonic point ($v_p=c_s$). The panels show (a) the density distribution (solid) and the distribution expected based on the mid-plane temperature (dashed), (b) the magnetic field components, (c) the radial velocity, (d) the poloidal (black) and toroidal (red) velocities with the corresponding Keplerian velocity ($R\Omega_K$; dashed blue line), (e) the vertical component of the Lorentz force relative to the gravitational force, and (f) plasma-$\beta$ for the total magnetic field strength (solid) and for the poloidal magnetic field strength (dashed).}
    \label{fig:path_0.3au}
\end{figure}

We start our discussion with the most laminar, intermediate radius region, Zone II. The location of this region drifts radially outward as the simulation progresses. By $t=1800 t_0$ (Fig.~\ref{fig:den_4panel}), it is located roughly between 0.1 and 0.5~au. The magnetic field line intersecting the midplane at a radius of 0.3~au is marked by a dashed line in Fig.~\ref{fig:den_4panel} (left panel). Fig.~\ref{fig:path_0.3au}(a) plots the density distribution along that field line, with the location of the sonic point marked. For comparison, we also plot the density profile expected for an isothermal (thin) disk with the temperature set to that at the midplane. Clearly, the presence of the magnetic field and the launching of a magnetized wind have significantly changed the vertical density profile of not only the disk itself, but also the region surrounding the disk. Specifically, the vertical density gradient is steeper near the disk surface compared to an isothermal (non-magnetic) profile. This difference is explained by a decreasing temperature along the path and by an increasing toroidal component of the magnetic field, $B_\phi$, which reaches a maximum near 0.02~au before decreasing again (Fig.~\ref{fig:path_0.3au}(b)). Indeed, the toroidal field becomes so strong near the disk surface that it dominates the thermal pressure (see Fig.~\ref{fig:path_0.3au}(f) and discussion below), and generates a downward magnetic pressure force that compresses the disk significantly. Above the disk surface, the magnetic pressure force has a positive radial component conducive to launching a wind.

The transition from disk to wind can be seen most clearly in the distribution of the radial component of the velocity $v_r$ along the field line (Fig.~\ref{fig:path_0.3au}(c)). The transition occurs approximately $\sim0.03$~au above the disk midplane. Material below this point moves radially inward with a speed of  $\sim$0.2~\kms, while material above this point is gradually accelerated outward. Beyond the sonic point the flow speed becomes more variable, because the wind has yet to reach a steady state. Even though the wind speed becomes supersonic at large distances, it remains well below the rotational speed except toward the edge of the simulation domain (see Fig.~\ref{fig:path_0.3au}(d)). In fact, $v_\phi$ is close to the speed needed to be rotationally supported against the gravitational pull from the central star in the cylindrical radial ($R$) direction (also shown in Fig.~\ref{fig:path_0.3au}(d) for comparison). The implication is that, just like the disk, the slowly expanding wind is almost entirely supported in the $R$-direction by rotation. The difference lies in the vertical direction; whereas the disk is supported against gravity (and magnetic compression) mostly by the thermal pressure gradient, the wind is supported mostly by the magnetic pressure gradient. This difference is highlighted in Fig.~\ref{fig:path_0.3au}(e), which shows that the ratio of the vertical component of the Lorentz force to the gravitational force changes sign at the disk surface/wind base and reaches unity just before the sonic point. The available evidence points to a magnetically dominated wind (Fig.~\ref{fig:path_0.3au}(f)) that is slowly lifted out of the (deep) gravitational well near the disk surface by the magnetic pressure gradient, consistent with the `magnetic tower' picture of \citet{1996MNRAS.279..389L}. 

Despite being much slower than the Keplerian speed at its footpoint, the wind launched from 0.3~au strongly affects the disk structure and dynamics. First, it rapidly depletes the disk material. To estimate a local disk mass depletion time, we define a magnetic flux tube containing an infinitesimally small magnetic flux $\Delta \Psi$ around the field line. The cross-sectional area of the flux tube is $\Delta A=\Delta \Psi /B_p$, where $B_p$ is the poloidal magnetic field. The rate of wind mass flux along the magnetic flux tube is
\begin{equation}
\Delta {\dot M}= \rho \Delta A \left(\frac{\bm{v}_p \cdot \bm{B}_p}{B_p}\right) = \rho \Delta\Psi \left(\frac{\bm{v}_p \cdot \bm{B}_p}{B_p^2}\right).
\end{equation}
This is to be compared with the amount of mass contained inside the flux tube
\begin{equation}
\Delta M = \int \rho \Delta A ds = \Delta \Psi \int (\rho/B_p) ds ,
\end{equation}
where the integration is along the field line, and is dominated by the mass inside the disk. The local disk mass depletion time due to wind mass loss is then
\begin{equation}
t_\text{dep}= \frac{\Delta M}{\Delta \dot{M}} = \int (\rho/B_p) ds  \left(\frac{\rho \bm{v}_p\cdot \bm{B}_p}{B_p^2}\right)^{-1}.
\end{equation}
In a steady state, the quantity~$\Delta \dot{M}$~should be constant along a field line. However, the wind does not reach a strict steady-state, and the non-steadiness introduces some variation to that combination, especially at large 
distances. We obtain a depletion time of $t_\text{dep}\approx 9.6$~yr, which is 58 times the local orbital period at the footpoint. This time scale is to be compared with the accretion time scale for the disk material, which is roughly
\begin{equation}
t_\text{acc}=r/v_r.
\end{equation}
For an average accretion speed of $\sim 0.2$~\kms~at 0.3~au, the accretion timescale is $t_\text{acc}\approx8.6$~yr, or 52 times the local orbital period. The fact that the accretion time is comparable to the depletion time means that a significant fraction of the disk will be ejected in the wind.

One way to quantify the disk accretion rate is through an effective $\alpha$ parameter, defined as
\begin{equation}
\alpha_\text{eff} = -v_r v_K/c_s^2 \label{eq:alpha},
\end{equation}
which, for $v_r\approx 0.2$~\kms, $v_K=54$~\kms, and $c_s=2.3$~\kms at 0.3~au, yields $\alpha_\text{eff}\approx 2$. This is much larger than the effective $\alpha$ typically obtained from turbulent MRI simulations ($\alpha\sim1/\beta$,~\citealt{2015SSRv..191..441H}), which highlights the dynamical importance of the slow disk wind launched even by a relatively weak initial poloidal magnetic field, in that it can still drive rapid disk accretion.

\subsubsection{Zone III: Avalanche accretion stream and slow midplane decretion}\label{sec:zoneIII}

\begin{figure}
    \begin{minipage}{0.5\linewidth}
	\includegraphics[width=1.0\columnwidth]{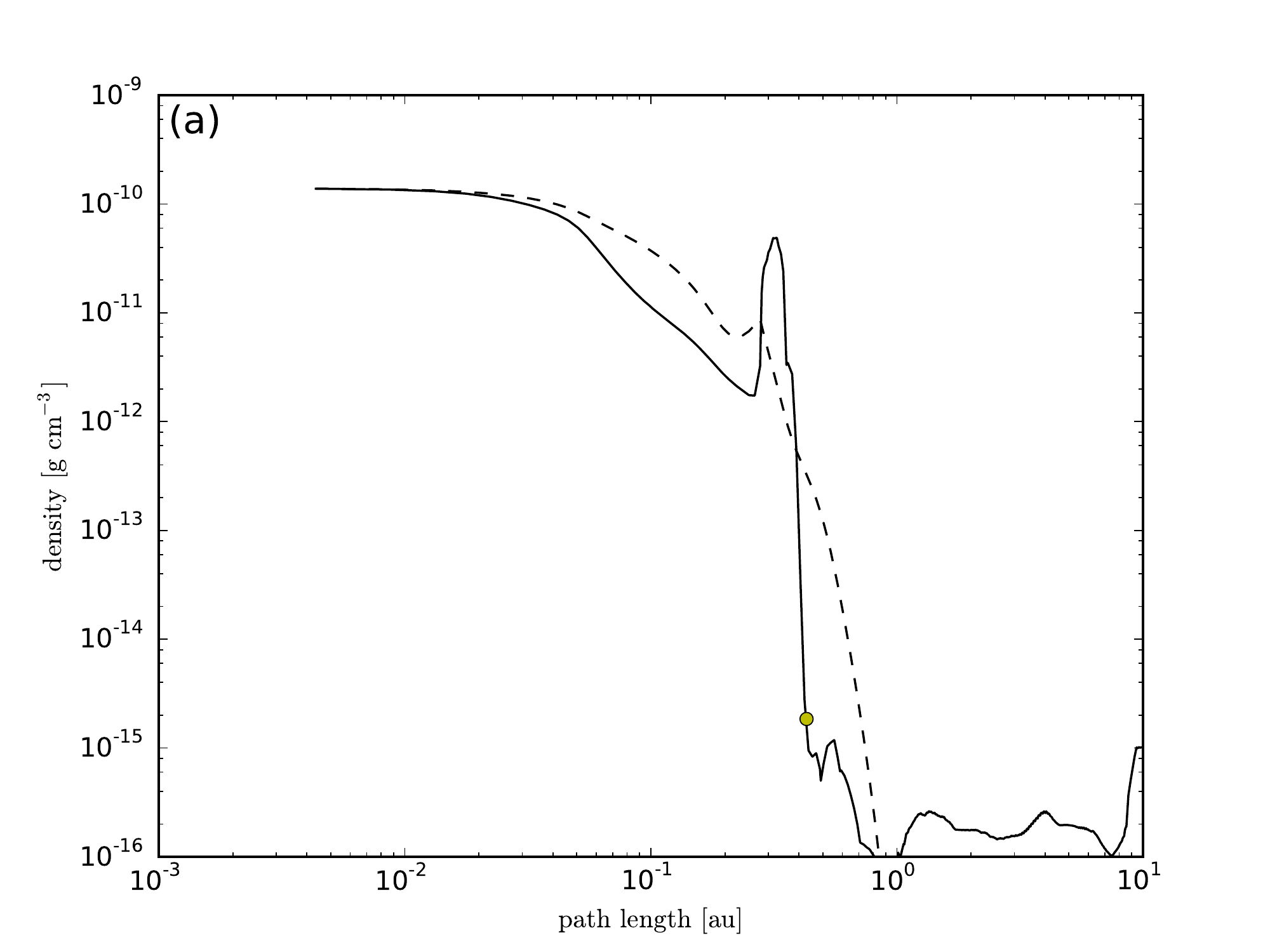}
	\end{minipage}
	\begin{minipage}{0.5\linewidth}
	\includegraphics[width=1.0\columnwidth]{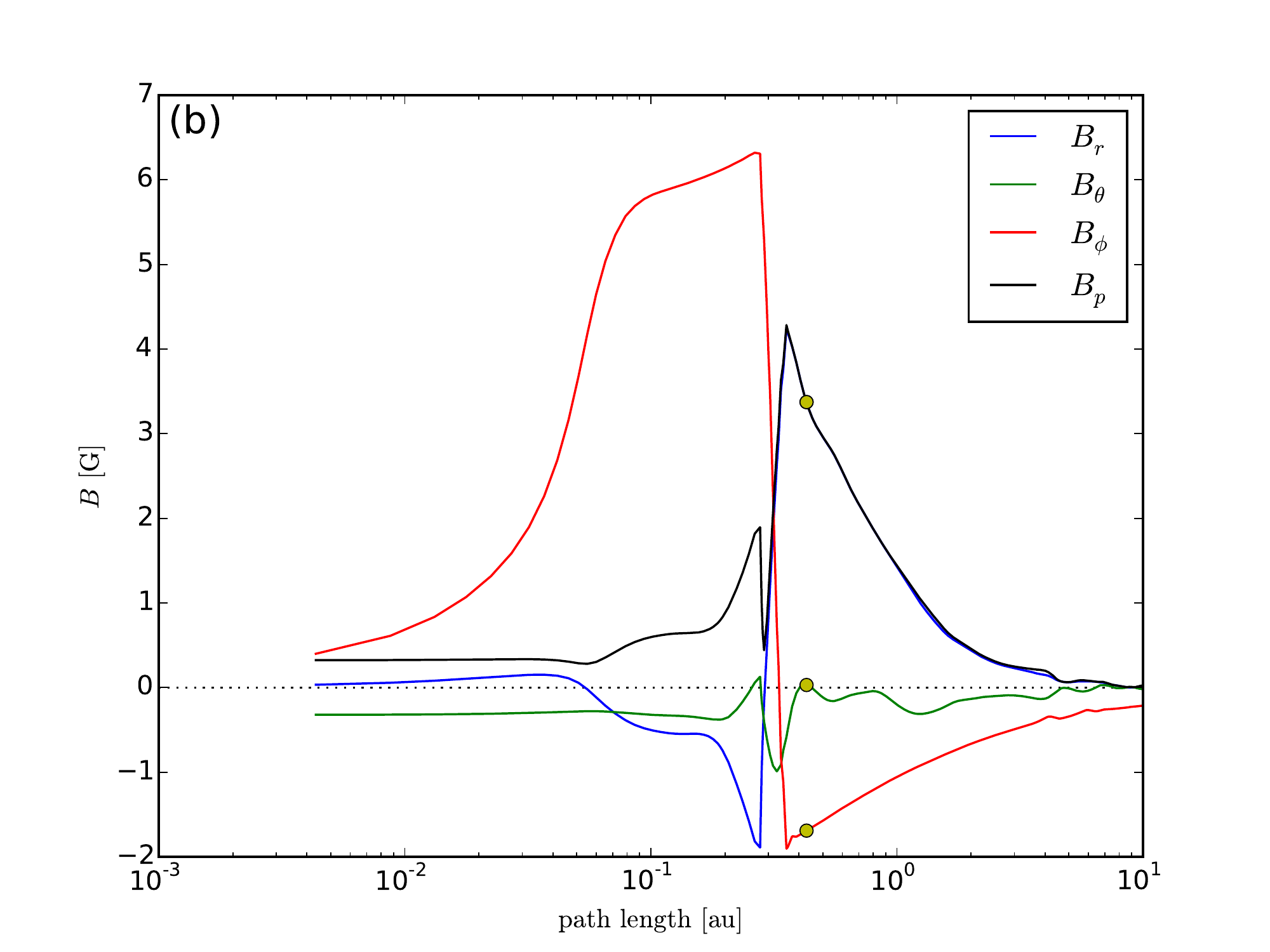}
	\end{minipage}
	\begin{minipage}{0.5\linewidth}
	\includegraphics[width=1.0\columnwidth]{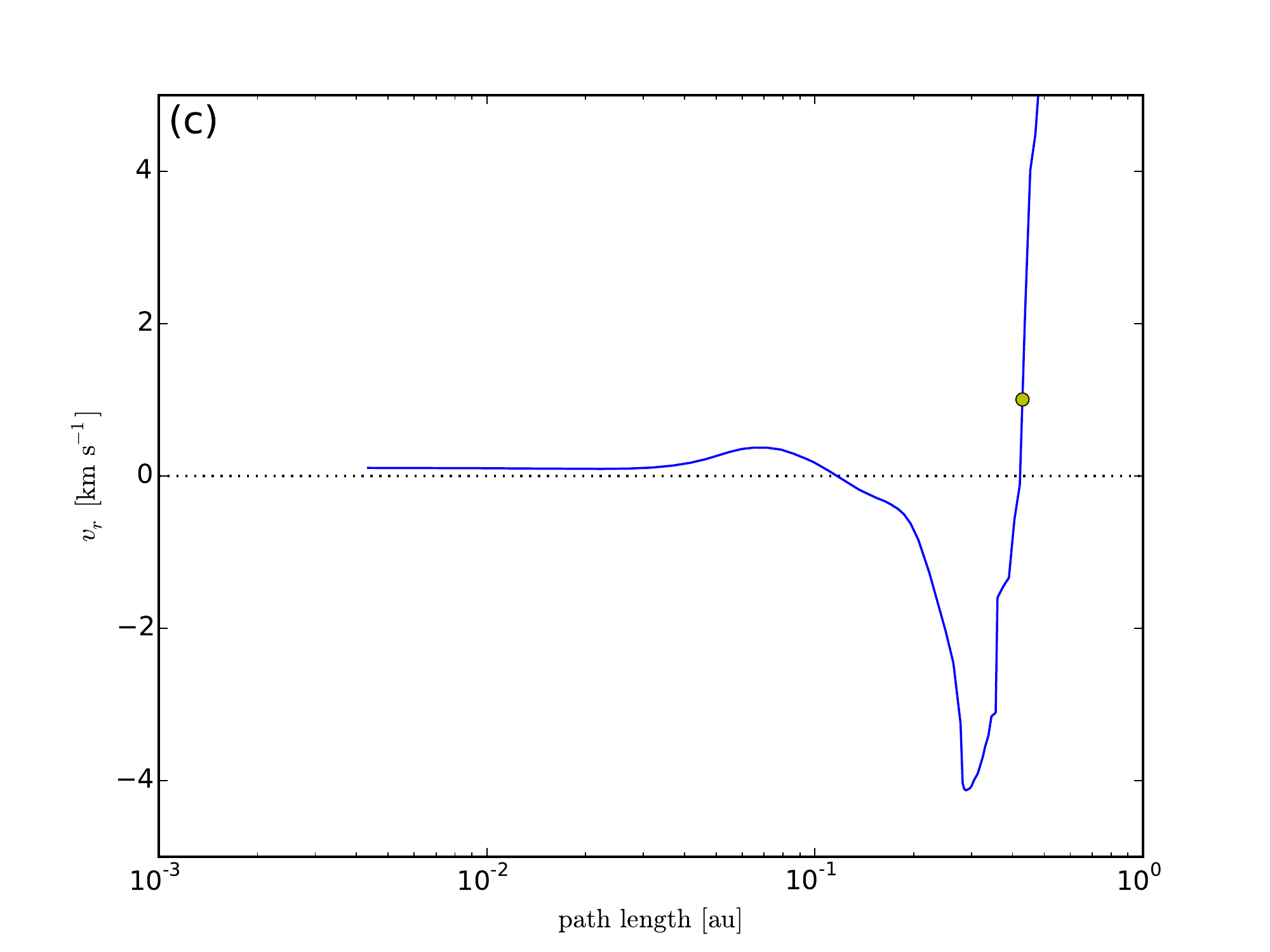}
	\end{minipage}
	\begin{minipage}{0.5\linewidth}
	\includegraphics[width=1.0\columnwidth]{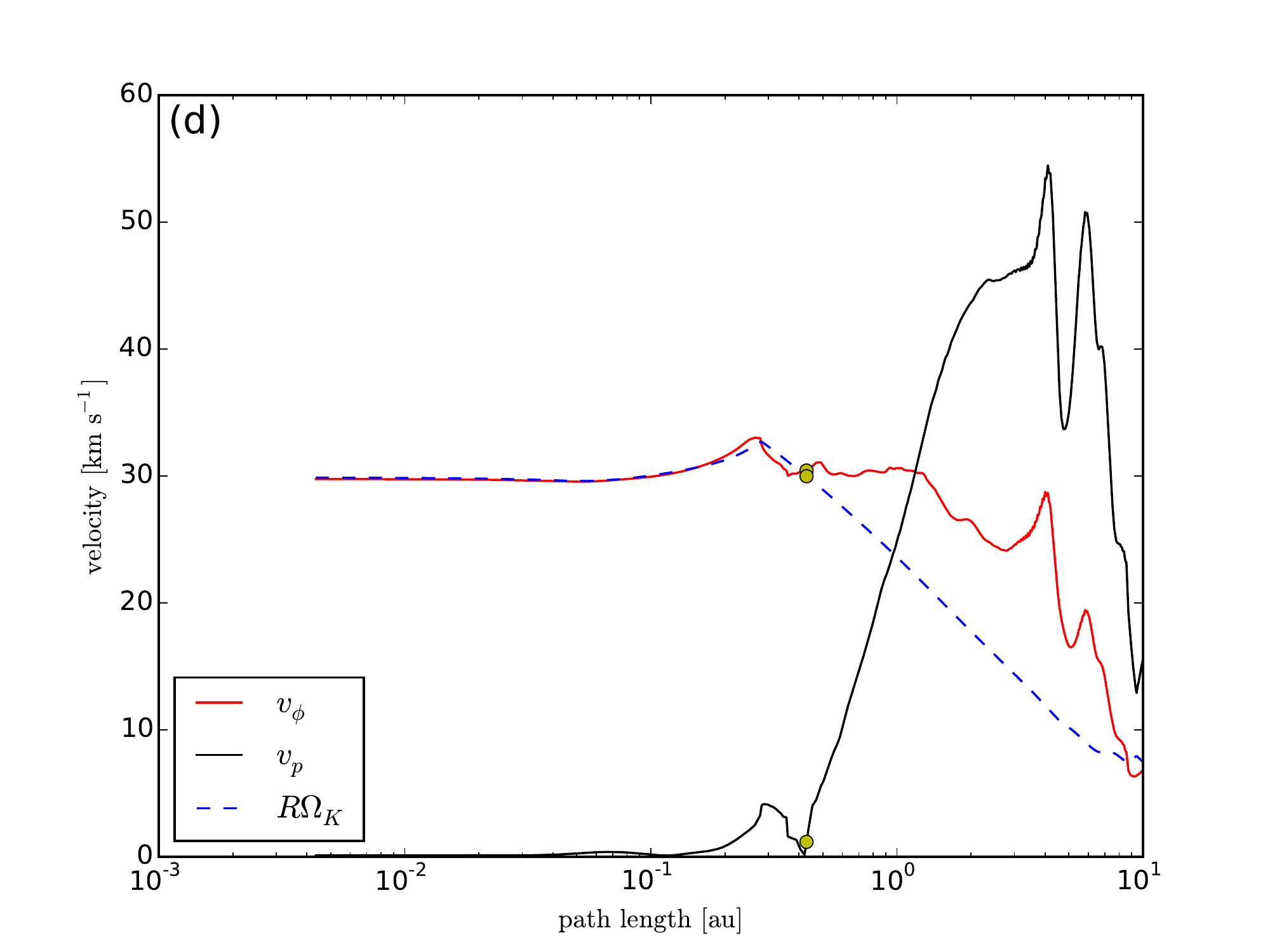}
	\end{minipage}
    \caption{Physical quantities plotted along a poloidal magnetic field line as a function of the distance along the line. The representative field line has a footpoint at $r=1$~au (Zone II) and can be seen in Fig.~\ref{fig:den_4panel} (white dashed line). Yellow circles show the sonic point in the outflowing wind ($v_p=c_s$). The panels show (a) the density (solid) and equilibrium density (dashed) distributions, (b) the magnetic field components, (c) the radial velocity, and (d) the poloidal (black) and toroidal (red) velocities with the corresponding Keplerian velocity ($R\Omega_K$; dashed blue line).}
    \label{fig:path_1au}
\end{figure}

For the outer disk region ($r>0.5$~au), we plot the physical properties of the gas and magnetic field along a representative field line that starts from a radius of $\sim 1$~au on the disk midplane (Fig.~\ref{fig:path_1au}). Panel (a) of Fig.~\ref{fig:path_1au} shows the density distribution as a function of the distance along the field line. We can see that near the midplane, the density profile is close to that expected for a (non-magnetic) isothermal disk (dashed line), indicating that the magnetic field is not important in the vertical force balance. Moving closer to the disk surface (where the magnetic pressure dominates the thermal pressure), magnetic compression causes the density to drop significantly below the isothermal value, as in Zone II. The difference is that the density in Zone III increases sharply again before dropping off precipitously. This density spike marks the `avalanche accretion stream' that is prominent in Fig.~\ref{fig:den_4panel}. It is the defining characteristic of Zone III, not only in the density structure but also in the magnetic field structure. 

From Fig.~\ref{fig:den_4panel}, it is clear that the dense stream occurs where the magnetic field line pinches severely, with a sharp reversal of the radial field component across it. This reversal shows up clearly in Fig.~\ref{fig:path_1au}(b), where all components of the magnetic field are plotted. The poloidal magnetic field line near the midplane first bows outward, with a small but positive radial component, as in Zone II (see Fig.~\ref{fig:path_1au}(b)). However, it is forced to bend sharply inward (see Fig.~\ref{fig:den_4panel}) by the rapidly infalling stream, producing a large (negative) value for $B_r$. The toroidal field component increases from the midplane towards the stream, dominating the other two field components before reversing direction across the stream (Fig.~\ref{fig:path_1au}(b)). 

As the material in the stream falls radially inward, it is orbiting faster relative to material both above and below it, as shown in Fig.~\ref{fig:path_1au}(d), where we plot $v_\phi$ along the representative field line. Simple geometric considerations show that the faster rotation at the tip of the sharply pinched field line will twist the inward-directing field into a toroidal field of positive sign and the outward-directing field into a toroidal field of the opposite sign, as shown in Fig.~\ref{fig:path_1au}(b). The twisted magnetic field geometry efficiently brakes the stream as it loses angular momentum to both the wind and the disk material that is magnetically linked to the stream at larger radii. Both $B_r$ and $B_\phi$ are amplified as the fast-rotating stream accretes, which leads to a stronger magnetic braking torque. The end result is a run-away collapse of the stream similar to the MRI, and hence the name `accretion avalanche' \citep{1998ApJ...508..186K}. At the time shown in Fig.~\ref{fig:path_1au}, the infall speed of the stream is supersonic and reaches about 2~\kms~(only the sonic point in the wind above the the stream is shown in the figure). The bulk of the disk material below the stream moves much more slowly and in the opposite direction, as shown in Fig.~\ref{fig:den_4panel} (left panel). The contrast between the fast accreting stream and slowly expanding disk is quantified in the radial velocity profile plotted in Fig.~\ref{fig:path_1au}(c). This `decretion' is caused by the disk material receiving angular momentum from the infalling stream and is opposite of the disk motion in Zone II (compare to Fig.~\ref{fig:path_0.3au}(c)).

Although the stream infalls supersonically, its infall speed is still much smaller than the rotation speed $v_\phi$ (Fig.~\ref{fig:path_1au}(d)). Indeed, the rotation speed is close to the value needed for the centrifugal force to balance the gravity in the cylindrically radial direction inside both the disk and the stream, indicating that both are close to being rotationally supported. A supersonic wind is launched above the stream, reaching speeds of $\sim$50~\kms, which is higher than that reached along the field line originating from 0.3~au (see Fig.~\ref{fig:path_0.3au}(d)). This speed difference is mainly due to lower mass loading, which allows the wind to experience greater acceleration despite rotating more slowly near the footpoint. Nevertheless, both the velocity and magnetic field are still dominated by the toroidal component over a large fraction of the wind, which is broadly similar to the wind driven from Zone II. The thermal pressure of the wind here is also completely dominated by the magnetic pressure, again similar to the wind from Zone II. Therefore, the formation of the stream appears to modify, but not completely disrupt, the wind launching and acceleration.

\subsubsection{Zone I: Rings and gaps in inner disk}\label{sec:zoneI}

\begin{figure}
	\includegraphics[width=1.0\columnwidth]{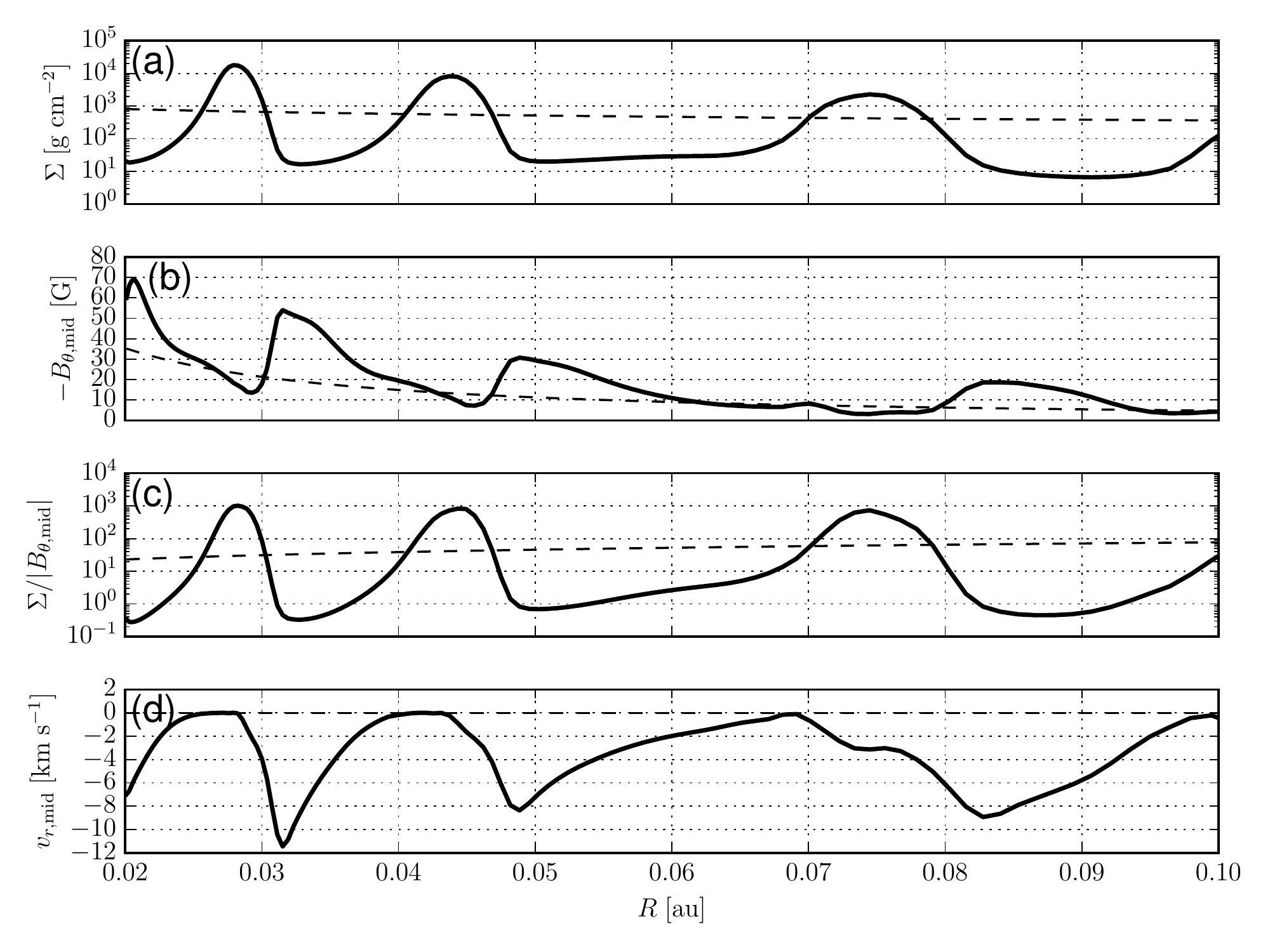}
	\caption{Ring and gap formation in Zone I of the reference simulation at $t=1250t_0$. (a) The surface density, (b) the vertical magnetic field strength at the midplane ($-B_{\theta,\rm{mid}}$), (c) the mass-to-flux ratio $\Sigma/\vert B_{\theta,\rm{mid}}\vert$ in units of g~cm$^{-2}$~G$^{-1}$, and (d) the radial velocity (negative means accretion towards the central source). The initial distribution of these quantities are shown for comparison (dashed lines).}
	\label{fig:mass2flux}
\end{figure}

The innermost disk region ($r<0.1$~au) is most strongly modified from its initial state during the simulation because it has the shortest rotation period and is located where the (poloidal) magnetic flux tends to accumulate through disk accretion. It is also the most variable region of the disk-wind system. The spatial variation of mass distribution is evident in Fig.~\ref{fig:den_4panel}, where the right panel shows a `face-on' view of the axisymmetric (one hemisphere) surface density normalized by the initial surface density distribution, $\Sigma_i=810~\gpercmsq~(r/r_0)^{-1/2}$, to highlight features at large radii. There is a striking contrast between the dense inner rings and their surrounding gaps; in fact, most of the mass of the inner disk is concentrated in these rings. Through the course of the simulation up to five rings are formed in Zone I. The first is formed just outside of the inner boundary and is quickly accreted after 200 inner orbital periods. The remaining rings all form in an inside-out manner; once one ring and gap are formed (with the gap located at a larger radius than the ring), another ring develops at a slightly larger radius than the first gap. Once formed, the multiple rings and gaps drift slowly inward maintaining their relative order.

Although the magnetic field topology near the rings and gaps is irregular, winds are still launched from Zone I. The winds that are launched from the gaps accelerate material to very high velocities ($v_r>100$~\kms), both because very little mass gets loaded onto the field lines anchored in the low-density gaps and because the vertical magnetic field strength peaks in the gaps (allowing for rigid rotation of the field lines out to a larger radius). This is quantified in Fig.\ref{fig:mass2flux}(a), where the column density as a function of radius at a representative time ($t=1250~t_0$) is plotted. One might naively expect the vertical component of the magnetic field to be concentrated in the high density rings, because the mass-to-flux ratio, $\Sigma/\vert B_{\theta,\rm{mid}}\vert$, would be conserved in the strict ideal MHD limit. However, there is a tendency for the vertical field to be \textit{weaker} in the dense rings than in the gaps. This anti-correlation is shown in Fig.~\ref{fig:mass2flux}(b), where $-B_{\theta,\rm{mid}}$ is plotted versus radius ($-B_{\theta,\rm{mid}}=\vert B_{\theta,\rm{mid}}\vert$ since the magnetic field points upward at the midplane). It is clear that the dense rings are less magnetized relative to both the initial disk value and the neighbouring gaps. The contrast between the rings and gaps is particularly striking for the distribution of the mass-to-flux ratio shown in Fig.~\ref{fig:mass2flux}(c).

Since the innermost rings are always observed to form first, it is natural to relate their formation to the inside-out development of the disk wind (i.e., at any given time, the disk wind is further developed at smaller disk radii because of their shorter dynamical times). The wind drains angular momentum from the disk, forcing the disk material to accrete and drag magnetic flux along with it. In the ideal MHD limit, the accumulation of mass at a given location would lead to a corresponding pile-up of magnetic flux at the same location. But in the resistive MHD of this simulation, the field lines diffuse away from the region of magnetic flux concentration. This leads to two key steps in the formation of a ring and gap. First, the magnetic flux begins to drop in the region where surface density starts to spike. The decreasing dynamical importance of the magnetic field in this region increases the accretion timescale of the dense ring, effectively creating a trap for mass accretion from larger radii. Second, the magnetic flux begins to accumulate just outside the growing overdensity. Now the low mass-to-flux region can efficiently drain disk material of angular momentum, quickly moving any remaining mass to smaller radii where it is added to the ring. The fast accretion of the material in the gaps onto the nearly stationary rings is shown in Fig.~\ref{fig:mass2flux}(d), where the radial component of the velocity at the disk mid-plane is plotted. This interplay of magnetic flux redistribution, angular momentum removal through magnetic wind, and disk accretion naturally leads to a configuration of a high mass-to-flux region (ring) just inside of a low mass-to-flux region (gap; Fig.~\ref{fig:mass2flux}(c)). Once an over-dense region is formed, its survival allows for the development of similar patterns at larger radii.

\subsection{Magnetic diffusivity and the disk-wind structure}

To summarize, there are three distinct zones that develop in the reference simulation. The dimensionless diffusivity parameter $D=\eta/(hc_s)$ decreases with radius as $D\propto r^{-1/2}$. Zone I is thus the most magnetically diffusive region initially, as measured by $D$. It is also the region where the most prominent rings and gaps form, because magnetic diffusion is needed for the redistribution of magnetic flux relative to the matter. In Zone II the magnetic diffusion is such that the outward diffusion balances the inward advection of the magnetic field, thereby establishing a laminar, quasi-steady disk-wind system. The least magnetically diffusive region, Zone III, develops a dense surface accretion stream that dominates the mass accretion in that region and drives the disk material below it to expand. These results suggest that the level of magnetic diffusion is a key factor in controlling the structure and dynamics of the coupled disk-wind system.

\section{Parameter study}\label{sec:param}

In order to determine how robust the basic features of the reference simulation are, especially the formation of rings, gaps and avalanche accretion streams, we performed several additional simulations in which we varied three dimensionless model parameters: $D_0$, $\beta$, and $\epsilon$. These characterize the disk magnetic diffusivity, field strength, and disk thickness (or temperature), respectively (see Table~\ref{tab:sims}). 

We find that all simulations show characteristics of at least one of the three Zones described in the reference simulation and all simulations develop variable winds (see Fig.~\ref{fig:mdot_comp}). Almost all simulations show enough spatial variation in the surface density that rings and gaps are readily apparent, and many of the simulations show dense avalanche accretion streams developing near the disk surface. In some cases the inner disks appear to be entirely dominated by rapid formation and break-up of short-lived accretion streams, forming a vertically extended `envelope' above the disk characterized by chaotic infall and outflow motions. In the following subsections we discuss the effects of each of the three dimensionless parameters in turn, focusing in particular on the lower diffusivity (D$\_$4), stronger field (model beta$\_$3), and higher temperature (t4, or thicker disk) cases (illustrated in Fig.~\ref{fig:edge_face} and Fig.~\ref{fig:mdot_comp}).

\begin{table}
  \centering
  \caption{Model parameters for all simulation runs. Note: the diffusivity parameters $D_0$ and $D_{m,0}$ are measured at the inner edge of the disk, and the simulation ref-x2grid has a higher resolution in the disk region (see Section~\ref{sec:grid}).}
  \label{tab:sims}
  \begin{tabular}{l c c c c c}
	\hline
    \hfill      & $\epsilon$ & $\beta$ & $\eta~[\rm{cm}^2~\rm{s}^{-1}]$ & $D_0$ & $D_{m,0}$\\
    \hline
    reference   & 0.05  & $10^3$            & $2.5\times10^{15}$    & 0.16 & 3.6  \\
    D\_4        & 0.05  & $10^3$            & $6.25\times10^{14}$   & 0.04 & 0.89 \\
    D4          & 0.05  & $10^3$            & $10^{16}$             & 0.64 & 14   \\
    beta\_3     & 0.05  & $3.33\times10^2$  & $2.5\times10^{15}$    & 0.16 & 2.1  \\
    beta3       & 0.05  & $3.0\times10^3$   & $2.5\times10^{15}$    & 0.16 & 6.2  \\
    beta100     & 0.05  & $10^5$            & $2.5\times10^{15}$    & 0.16 & 36   \\
    t4          & 0.1   & $10^3$            & $10^{16}$             & 0.16 & 3.6  \\
    ref-x2grid  & 0.05  & $10^3$            & $2.5\times10^{15}$    & 0.16 & 3.6  \\
    \hline
  \end{tabular}
\end{table}

\subsection{Resistivity}\label{sec:resistivity}

\begin{figure}
    \centering
	\includegraphics[width=0.95\columnwidth]{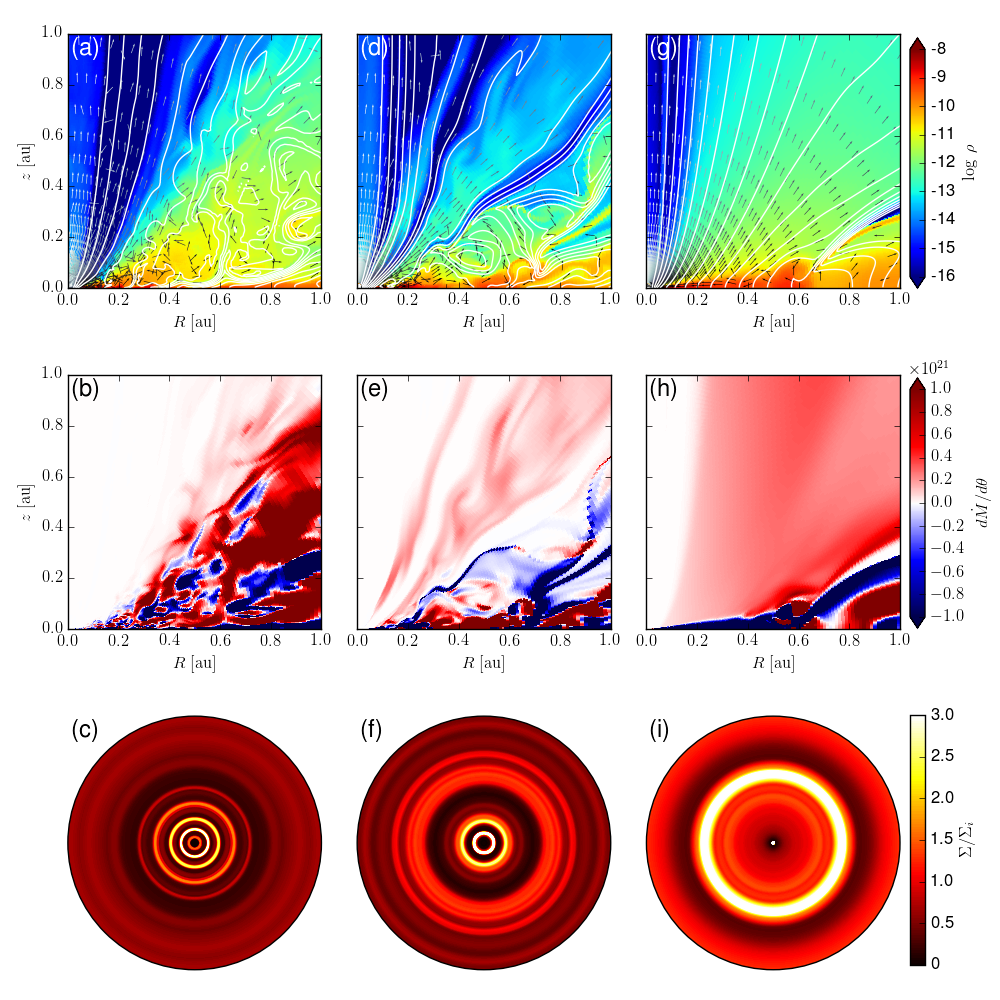}
    \caption{Snapshots of three simulations at $t=1650t_0$. The left, middle, and right columns correspond to simulations D\_4, beta\_3, and t4, respectively. The top row shows the mass density (g~cm$^{-3}$) in logarithmically spaced colour contours with magnetic field lines in white and velocity (unit) vectors in greyscale. The middle row shows the radial mass flux per unit polar angle $d\dot{M}/d\theta=2\pi r^2\rho v_r \sin{\theta}$ where negative (blue) values correspond to infall and positive (red) to outflow. The bottom row shows the `face-on' view of the axisymmetric surface density normalized to its initial distribution for $r\leq1$~au. (See the supplementary material in the online journal for animated versions of this figure, including the reference simulation.)}
    \label{fig:edge_face}
\end{figure}

We start with model D\_4, where all parameters are the same as in the reference case except for the (constant) resistivity $\eta$, which is reduced by a factor of four (thereby reducing $D_0$ by the same amount). The reduction in $\eta$ makes the magnetic field lines better coupled to the material in the disk. We expect this to facilitate the formation of avalanche accretion streams, as in Zone III of the reference simulation, and this is indeed the case. Movies of simulation D\_4 (see the supplementary material in the online journal for animated versions of Fig.~\ref{fig:edge_face}) show that avalanche accretion streams form continuously, starting as early as $\sim 10$~inner orbital periods. Many of the streams are pushed up high into the disk corona by the growing toroidal magnetic fields beneath them (as in the case of the stream in Zone III of the reference simulation, see Fig.~\ref{fig:path_1au}(b)). They are eventually disrupted as they fall radially inward and towards the midplane. The constant formation and disruption of the (often elevated) avalanche accretion streams leads to a thick, clumpy envelope above the disk that is highly inhomogeneous in density, velocity, and magnetic field. This stream-produced envelope can be seen clearly in Fig.~\ref{fig:edge_face}(a), where its maximum vertical extent is comparable to the cylindrical radius and is positioned between the denser equatorial disk and the more tenuous polar wind region. While the (poloidal) field lines in the polar wind region are well organized, those in the envelope are disordered with severe bunching of field lines in some places and looping (due to reconnection) in others. The disordered field is a reflection of the chaotic motions inside the envelope characterized by simultaneous infall and outflow. The motion of the envelope is best seen in Fig.~\ref{fig:edge_face}(b), where the spatial distribution of mass flux per unit polar angle, $d\dot{M}/d\theta=2\pi r^2\rho v_r \sin{\theta} $, is plotted. The mass accretion in the envelope is dominated by the fragments of disrupted avalanche accretion streams (blue in middle row of Fig.~\ref{fig:edge_face}). They are mixed together with packets of outflow (red) that sometimes extend to the disk midplane. The picture is reminiscent of the MRI-driven channel flows that are seen, e.g., in the thick disk simulations of \citet{2002PASJ...54..121K}, and they share the same physical origin -- the exchange of angular momentum between magnetically connected material located at different radii. In our thin-disk case, the difference is that the alternating pattern of inflow and outflow occurs mostly in the envelope above the disk rather than inside the disk. This dynamic envelope is a new feature that formally does not exist in the reference simulation, although it is intimately related to the avalanche accreting stream of Zone III. 

\begin{figure}
    \centering
	\includegraphics[width=0.75\columnwidth]{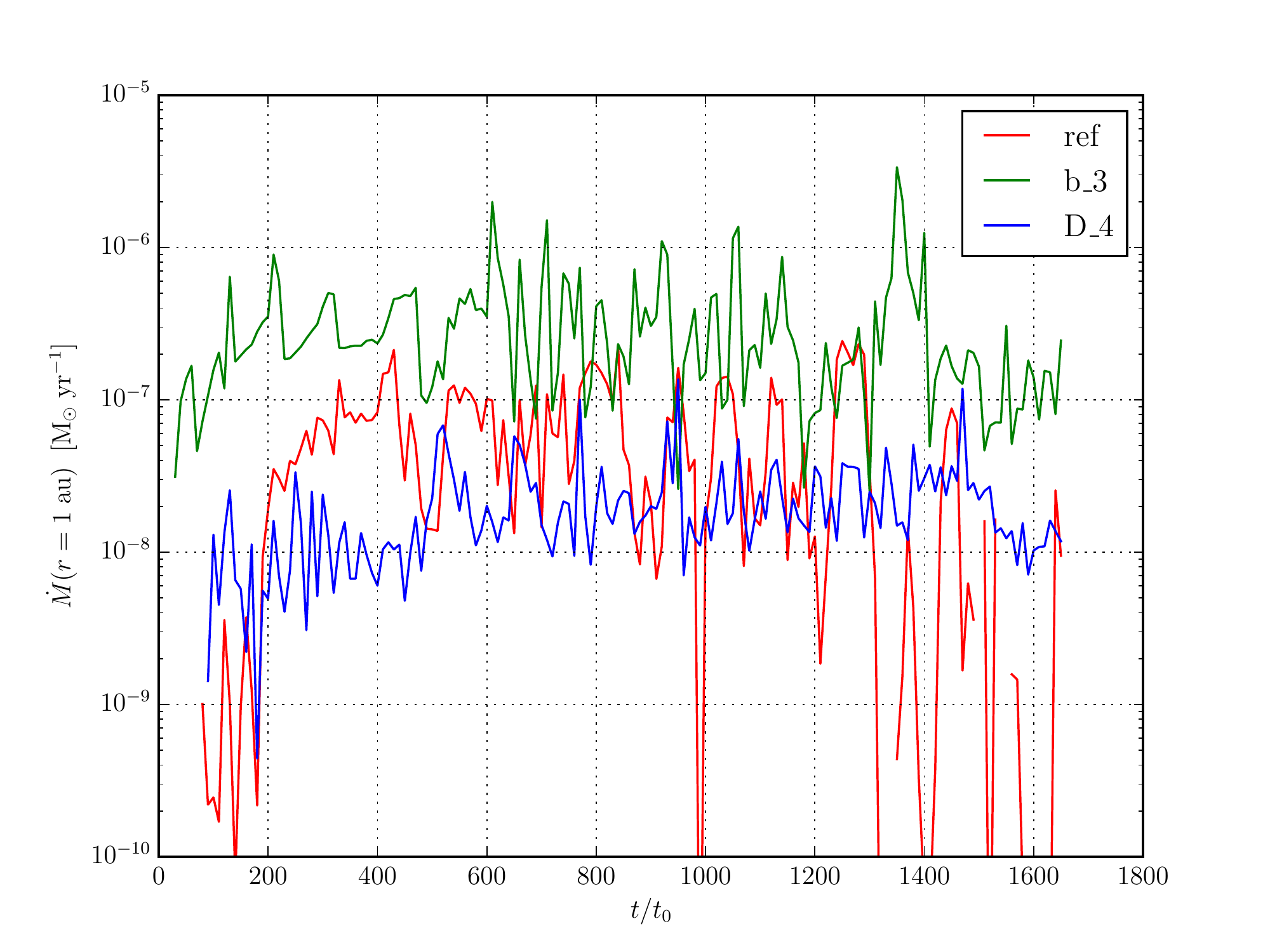}
    \caption{Mass outflow rate (\msunperyr) through a hemisphere of $r=1$~au as a function of time for three simulations (reference, beta\_3, and D\_4). Only the mass outflow rate of the fast velocity component ($v_r>100$~\kms) is shown.}
    \label{fig:mdot_comp}
\end{figure}

In addition to the envelope, a wind is also launched in the low-resistivity case (model D\_4; see also \citealt{2009ApJ...691L..49S,2013A&A...552A..71F}). The wind can be seen in Fig.~\ref{fig:edge_face}(a), especially in (but not limited to) the polar region. The wind is highly variable, as illustrated in Fig.~\ref{fig:mdot_comp} (blue line), where the mass loss rate for the fastest wind component ($v_r>100$~km/s) is plotted as a function of time. The mass loss rate is comparable to that in the reference case, especially in the second half of the simulation. The highly variable wind and the chaotic envelope create radial structures in the disk. Fig.~\ref{fig:edge_face}(c) shows a face-on view of the surface density normalized by its initial distribution. Several rings and gaps are clearly visible, and they are quantified in Fig.~\ref{fig:surfden_panels}(b), where the surface density is plotted as a function of radius. The development of rings and gaps starts at small radii, and gradually spreads to large radii, because the dynamical time increases with radius. Some of the rings appear to change quickly, while others are more stationary. These features are present throughout the simulation, suggesting that they are a robust characteristic of the coupled disk-envelope-wind system. We find that the surface density anti-correlates with the poloidal field strength on the disk, with the dense rings typically less magnetized than the gaps, as in Zone I of the reference simulation. This indicates that magnetic flux redistribution through either the relatively small resistivity or turbulent reconnection, is playing a role in creating the rings and gaps. Since the disk is tightly connected to the stream-dominated envelope, the streams are also expected to play an important role in gap and ring formation. This role is difficult to quantify precisely, however, because of the chaotic flow pattern in the disk and envelope. A cleaner case of stream-induced ring formation will be presented in the higher-temperature case below (see Section~\ref{sec:temp}). We have also carried out an ideal MHD simulation with the same parameters except for $\eta=0$, and found results broadly similar to this low-resistivity case.

In contrast to D\_4, the more resistive model D4 is much more laminar. A wind is launched steadily throughout the run over most of the disk radii that have had time for at least one orbit to take place, as in the smooth Zone II of the reference run. There are no avalanche accretion streams in the simulation. A new feature of model D4 is that, because of its high resistivity, the wind from the innermost part of the disk becomes weaker with time as the magnetic flux in the region diffuses to larger radii. By $t=1800 t_0$, the region of weakened wind launching extends to a radius $\sim 0.2$~au (Fig.~\ref{fig:surfden_panels}(e)). In this region, the rate of angular momentum removal by the wind is significantly reduced, allowing mass to accumulate there. This results in a large plateau in the surface density profile followed immediately by a gap at 0.2~au, with a drop in surface density of one order of magnitude. Model D4 therefore provides another example of gap formation through a wind of varying strength at different radii, in addition to Zone I of the reference simulation.
At 2200 inner orbital periods, the mass in the disk within $r<0.2$~au has doubled, and the magnetic flux has dropped well below its initial value. The magnetic field remaining in the dense plateau is too weak to break the surface density into rings and gaps, unlike Zone I of the reference case. The drastically different behaviors of models D\_4, D4, and the reference model underscores the key role of the magnetic diffusivity in determining the structure and dynamics of the coupled disk-wind system. 

\subsection{Magnetic field strength}\label{sec:bfield}

We will focus first on model beta\_3, where the (vertical) magnetic field is $\sqrt{3}$ times stronger than that of the reference case at the disk midplane (i.e., $\beta$ is decreased by a factor of three). One expects a stronger magnetic field to drive a more powerful disk wind, and this is indeed the case. Fig.~\ref{fig:mdot_comp} shows that the wind mass loss rate is typically well above that of the reference case. It is also clear from Fig.~\ref{fig:mdot_comp} that the disk wind in this case is just as variable as, if not more variable than, that in the reference case.

The flow pattern in and around the disk in model beta\_3 is more similar to that in the least resistive model D\_4 than to the reference case. Specifically, it is dominated by the constant formation and disruption of dense avalanche accretion streams (like that seen in Zone III of the reference case); the quasi-laminar region (Zone II) has all but disappeared. These changes are expected because the magnetic torque exerted by a stronger field removes angular momentum from the disk more efficiently, allowing the accreting streams to develop earlier and at smaller radii. The first prominent stream develops within 150 inner orbital periods (earlier than the reference run but later than model D\_4) and terminates near $r=0.2$~au. By 500 inner orbital periods the inner disk is almost completely restructured by the interaction of multiple avalanche streams, as in model D\_4, again leading to the formation of a thick envelope where matter moves rather chaotically both inward and outward (see Fig.~\ref{fig:edge_face}(d)). Compared with model D\_4 at similar times, the number of avalanche streams is smaller, the stream-dominated envelope is thinner (typically reaching only $z\sim 0.5 R$ compared to $\sim R$), and the wind above the envelope is more massive (compare Fig.~\ref{fig:edge_face}(e) and (b)). In addition, the expansion of disk material near the midplane is more clearly visible in model beta\_3; for example, about half of the disk inside 0.6~au is decreting at the time shown in Fig.~\ref{fig:edge_face}(e). Nevertheless, the flow patterns in the two cases are fundamentally similar: both are dominated by the avalanche accretion streams, as they are not suppressed by a large enough magnetic diffusivity. We will return to this important point toward the end of the subsection.

\begin{figure}
    \centering
    \includegraphics[width=0.45\columnwidth]{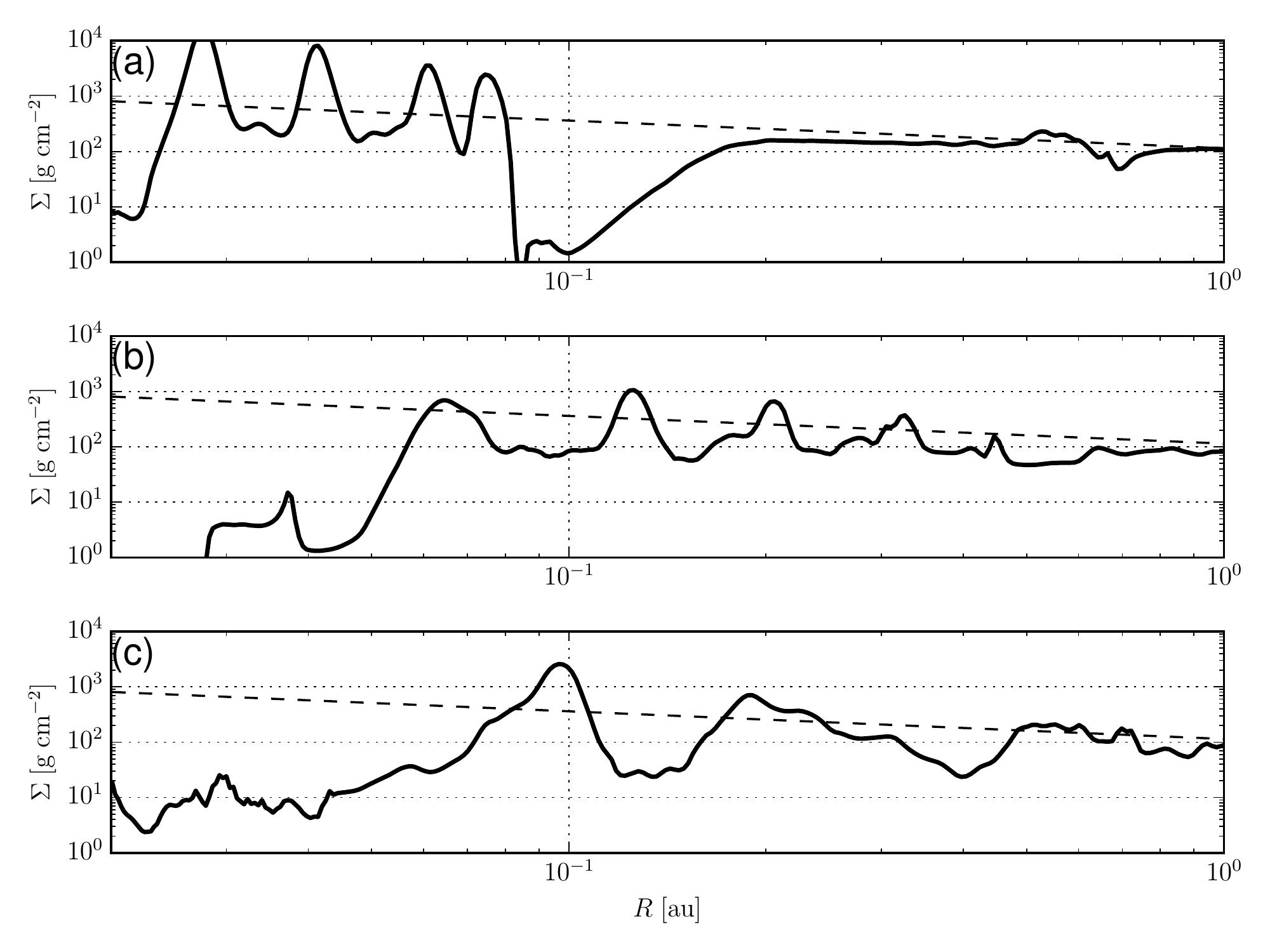}
	\includegraphics[width=0.45\columnwidth]{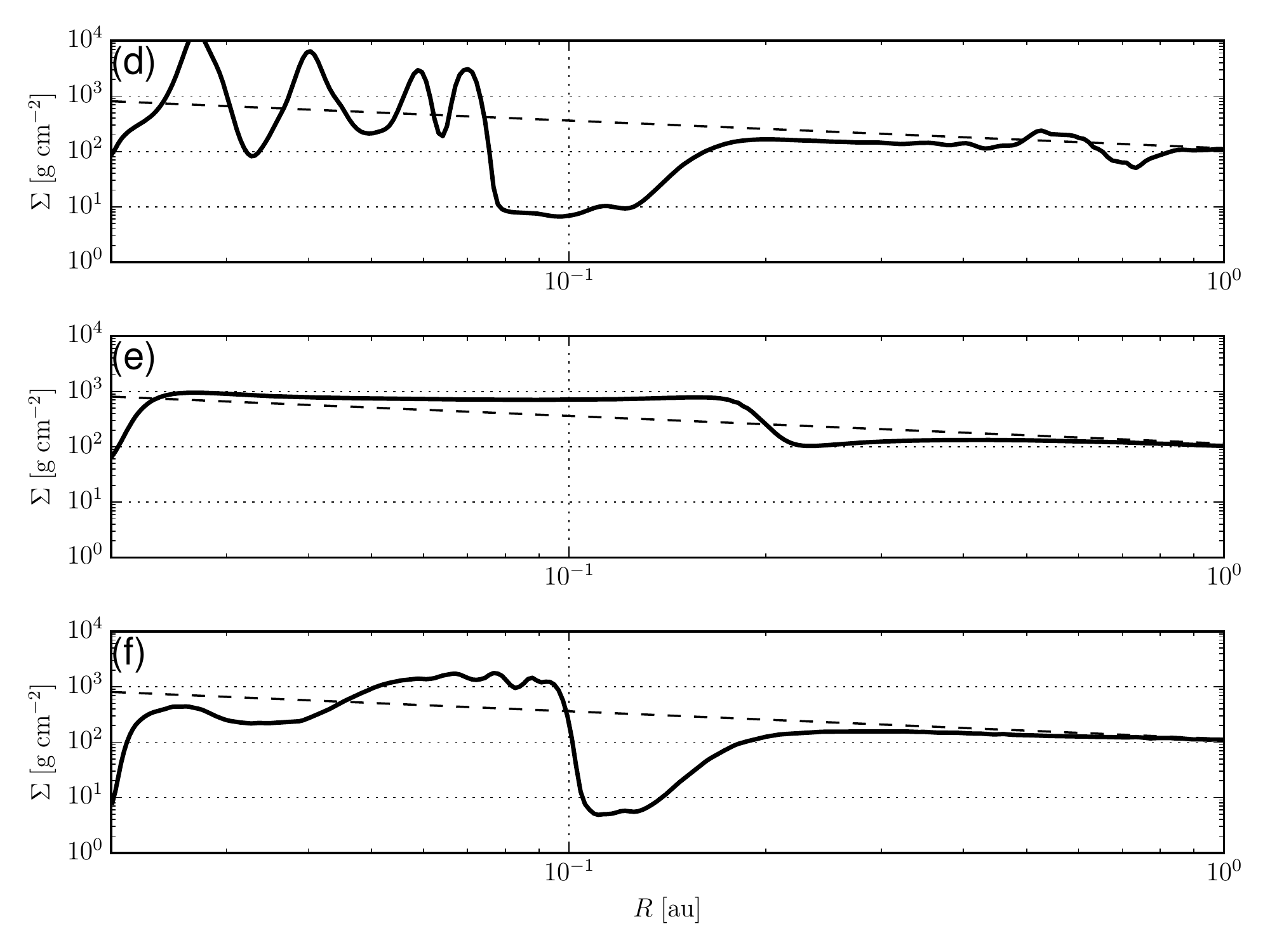}
    \caption{Surface density profiles. Both columns compare the reference simulation (top row) to two simulations below it at a given simulation time. The initial surface density profile is shown for comparison (dashed). Left: (a) reference, (b) D\_4, and (c) beta\_3  at $t=1650t_0$. Right: (d) reference, (e) D4, and (f) beta3 at $t=1800t_0$.}
    \label{fig:surfden_panels}
\end{figure}

The strong, highly variable wind and the stream-dominated envelope create structures in the disk surface density distribution, which can be seen in Fig.~\ref{fig:edge_face}(f) and Fig.~\ref{fig:surfden_panels}(c). There are numerous rings and gaps at the time shown ($t=1650t_0$). Their development starts near the inner boundary and spreads outward with time, as the disk material at an increasingly larger radius becomes affected by the magnetic wind launching and the formation of streams. We find an anti-correlation between the surface density and vertical magnetic field strength in the midplane (as in model D\_4 and Zone I of the reference case), although the interpretation of the anti-correlation is again complicated by the chaotic motions in the disk and envelope.

The weaker field model, beta3, is dominated by a quasi-laminar region beyond 0.1~au (similar to Zone II of the reference run), and lacks a region with avalanche streams. In fact, not a single stream is formed throughout the duration of the simulation. Within a radius of 0.1~au, a wind appears only intermittently and is less efficient in carrying away angular momentum compared to the wind launched beyond this radius. This allows mass accumulation at small radii, forming the high surface density plateau, as in model D4 (Fig.~\ref{fig:surfden_panels}(f)). This plateau is adjacent to a prominent gap at $r=0.1$~au, which separates the variable inner and steady outer wind launching region, similar to the boundary between Zone I and Zone II in the reference case. This difference is further illustrated by the weakest magnetic field model, beta100. In this case, an outflow is still launched, but is unable to escape the simulation domain; instead, it inflates a bubble up to $z\sim1$~au that is dominated by the toroidal magnetic field with $\beta\approx5\times10^{-2}$. The disk remains laminar and mostly unchanged from its initial state.

Taken together, the four models with different field strengths but the same resistivity (beta100, beta3, reference, and beta\_3) show a clear trend: as the field becomes stronger, the disk-wind system becomes less laminar, with the avalanche accretion streams becoming increasingly more important. This is the same trend that we see in Section~\ref{sec:resistivity}, where the importance of the avalanche accretion streams increases as the resistivity decreases for a given field strength. The similarity indicates a deep connection between the effects of a stronger magnetic field and a lower resistivity. 

The connection can be understood with the help of a second diffusivity parameter $D_m \equiv \eta/(hv_A)$, where $v_A$ is the Alfv\'en speed; it is the inverse of the standard Lundquist number in plasma physics, $S\equiv h v_A/\eta$, and is related to the first diffusivity parameter as $D_m=\sqrt{\beta/2}\ D$. It measures the importance of the resistive magnetic diffusion on magnetically induced motions. Obviously, a lower resistivity would lead to a better coupling between the magnetic field and the matter. Perhaps less obvious is that a stronger magnetic field would also lead to a better field-matter coupling. This is because a stronger field induces a faster motion over a given length-scale, which leaves less time for the field lines to diffuse resistively relative to the matter. In this sense, the diffusivity parameter $D_m$ provides a better indicator for how the coupled disk-wind system behaves: the smaller $D_m$ is, the more important the avalanche accretion streams would become. Conversely, the streams are suppressed by a large enough $D_m$ (analogous to the suppression of the MRI). This explanation applies not only to models with different parameters, but also across different regions of the reference model. For example, it naturally explains why an avalanche accretion stream develops only at the largest radius (Zone III) of the reference simulation, where the parameter $D_m$ is the lowest because it initially decreases with radius as $r^{-1/2}$. 

\subsection{Disk thickness/temperature}\label{sec:temp}

In the simulation t4, the disk aspect ratio $\epsilon$ is increased by a factor of two, thereby increasing the initial disk temperature by a factor of four (see Table~\ref{tab:sims}). To keep $\beta$ and $D_0$ (and $D_{m,0}$) the same as the reference run, we increase the initial poloidal field strength $B_{p,0}$ by a factor of two and the resistivity $\eta$ by a factor of four. In this simulation, there is an extended period of time where a stable disk wind is launched within a radius of $r<0.5$~au. Immediately outside of this is an avalanche accretion stream that terminates at $r\approx0.5$~au. The mass deposited by the stream grows fast enough to form a ring (see Fig.~\ref{fig:edge_face}(g) and (i)), while the disk region beneath the stream moves radially outward (Fig.~\ref{fig:edge_face}(h)). This provides the cleanest evidence that a ring can be formed directly from a stream. This ring formation process was also present in the reference model, where it increased the surface density at $r=0.5$~au by a factor of five relative to its neighbouring gap (at a slightly larger radius). The effect is even more noticeable in model t4, with the surface density contrast between the ring and gap increased to a factor of 40. We again see an anti-correlation between the surface density and magnetic flux for the rings and gaps created via an avalanche accretion stream. Naturally, material being deposited at the end of the accretion stream will increase the disk mass locally, but the magnetic flux will not increase here because the mass accretion through the stream is nearly parallel to the field that confines the flow. The magnetic flux is reduced in the region under the stream that has been moving radially outward.

\subsection{Resolution}

In order to study the effects of the grid resolution on the simulations, we increase the disk resolution by a factor of two in the $\theta$-coordinate direction in model ref-x2grid. The grid now contains 240 uniformly-spaced cells from $\theta=\pi/3$ to $\pi/2$ and 120 non-uniform cells from $\theta=\pi/3$ to the polar axis ($\theta=0$) with a constant ratio for the widths of adjacent cells (see Section~\ref{sec:grid}). In order to match the cell aspect ratio with all previous simulations, we also decrease the value of $dr_0$ by two ($dr_0$ is the width of the first cell on the $r$-grid). This simulation has the same qualitative evolution as the reference run, and we can separate the disk evolution into three zones with radial delineations matching those described for the reference run. The primary difference seen between the reference run and the high resolution run is in the number of rings/gaps formed in Zone I. There are up to three rings at any one time in the high resolution run, but by $t=1650t_0$ they have accreted to form one large ring outside the inner boundary at $r=0.03$~au. The remaining zones follow similar behavior to the reference run, indicating that the qualitative behavior of the disk-wind system -- especially the disk-wind variability and the ring, gap, and stream formation -- are independent of the simulation resolution.

\section{Discussion}\label{sec:dis}

\subsection{Comparison with other work}

\subsubsection{Steady disk wind}

Although one of the focuses of this work is on the variability of winds and their impact on forming radial structures in disks, previous studies have focused largely on steady winds such as those seen in Zone II of the reference simulation. Many authors aim to explore the parameter space of disk properties in an effort to connect them to properties of the wind (e.g., \citealt{2009MNRAS.400..820T,2010A&A...512A..82M,2012ApJ...757...65S,2014ApJ...793...31S}; \citetalias{2016ApJ...825...14S}). In these works, the resistivity often takes the form $\eta\propto h v_A$ inside the disk and zero outside. The scaling of this expression is chosen such that the inward accretion and outward diffusion of magnetic flux are able to reach a quasi-steady state, with a fraction of the disk material getting launched into the wind, thereby driving disk evolution through angular momentum removal. Such steady-state solutions, while important for illuminating the physics of wind-launching and for making connection with analytic work, are not guaranteed in nature. There is no \textit{apriori} reason why the Lundquist number $S\equiv \eta/(h v_A)$ should be spatially constant or of order unity, especially in the inner disks where the electron number density varies rapidly with radius (caused by collisional ionization of the alkali metals and/or thermionic emission from dust grains; \citealt{1981PASJ...33..617U,2015ApJ...811..156D}). We have shown that, in the simplest illustrative case of a constant resistivity, a steady-state is not reached in general.

\citetalias{2016ApJ...825...14S}~have shown that winds can be launched for disks with plasma-$\beta$ from $10^{0.7}$ to $10^{3.5}$. They have also confirmed that there is a critical magnetization below which the launching mechanism transitions from the classical magnetocentrifugal mechanism of \citet{1982MNRAS.199..883B} to the magnetic tower mechanism driven by the toroidal magnetic pressure gradient \citep{1996MNRAS.279..389L}. This transition occurs near $\beta\approx30$. Our simulations typically have $\beta\approx10^3$ (specifically in Zone II of the reference simulation, $\beta=10^{2.36}$ at $r=0.3$~au after $t=1800t_0$), so our finding that the wind is in the magnetic tower regime (see Section~\ref{sec:zoneII}) is consistent with previous work. This is also consistent with the 1D analytic model of \citet{2016ApJ...818..152B}, who found that the disk wind tends to be driven by a magnetic pressure gradient unless the Alfv\'en speed near the disk surface is much larger than the local Keplerian speed. Given the small magnetic lever arm in this regime, the mass loading rate is high and the wind speed remains low in the steady part of the wind.

\subsubsection{Rings and gaps}\label{sec:ring_gap}

In the era of ALMA, rings and gaps are being observed in an increasing number of disks, including, e.g., HL Tau \citep{2015ApJ...808L...3A}, TW Hya \citep{2016ApJ...820L..40A}, and HD 163269 \citep{2016PhRvL.117y1101I}. The most commonly invoked explanations for such features are that they are cleared by planets \citep{2015ApJ...809...93D} or are the result of the condensation of abundant volatile compounds (i.e., snowlines; \citealt{2016ApJ...818L..16Z}). In almost all of the disks simulated in this work, gaps are opened with surface density contrast ratios of at least one order of magnitude. The fact that gaps can be created purely through MHD processes is interesting in its simplicity. The mechanism requires only a magnetized disk with an accretion rate that varies as a function of radius. In our simulations, one way to achieve this variation is through an MHD disk wind that carries away differing amounts of angular momentum from different radial locations in the disk. The magnetic torque acting on the surface of the disk then results in a mass accretion rate of $\dot{M}_\text{acc}(R) \approx R B_\theta B_\phi/\Omega_K$ (\citealt{2016ApJ...818..152B}, see their equation 19). Therefore, a magnetic disk wind can open a gap (or create a ring) so long as $B_\theta B_\phi$ reaches a local maximum (or minimum) somewhere in the disk. In our particular problem setup, the rings and gaps are formed within a few au of the central mass because only such inner regions have short enough orbital periods to evolve through multiple orbits in the duration of the simulation. However, this mechanism should operate at larger disk radii as long as the magnetic disk wind is able to redistribute angular momentum on these scales as well \citep{2016Natur.540..406B}. 

Several other MHD mechanisms have been proposed for the creation of radial surface density maxima in disks. For example, the simulations of \citet{2015A&A...574A..68F} and \citet{2016A&A...590A..17R} have shown that a surface density bump is formed inside the MRI dead zone of a disk, while a gap is opened up outside the dead zone due to MRI-driven mass accretion. The authors find an anti-correlation between the magnetic field and surface density (similar to that shown in Fig.~\ref{fig:mass2flux}) as the magnetic field accumulates in the MRI-active gap outside the dead zone. A related phenomenon are the so-called `zonal flows' \citep{2009ApJ...697.1269J}. Zonal flows occur when a radial pressure gradient is balanced by the Coriolis force, leading to alternating radial bands of sub and super-Keplerian flows (see \citealt{2015arXiv150906382A} for a recent review). These flows develop in MRI disk simulations from large-scale variations in the Maxwell stress ($B_rB_\phi$) and result in an anti-correlation between the magnetic pressure and the mass density \citep{2009ApJ...697.1269J,2013MNRAS.434.2295K,2014ApJ...796...31B}. In global disk simulations, zonal flows and zonal (magnetic) fields can be generated by the Hall effect because the additional Hall term in the induction equation can act to radially confine the vertical magnetic flux \citep{2016A&A...589A..87B}. The rings and gaps that form in our simulations share some characteristics with the zonal flows (e.g., both show an anti-correlation between surface density and magnetic pressure), but their formation mechanisms are quite distinct.

In addition to variable magnetic winds, our simulations show that rings can also form directly out of rapidly accreting streams. This ring formation mechanism is observed most clearly in the simulation with a larger initial disk thickness (or temperature;  model t4, Section~\ref{sec:temp}), and to a lesser extent in the reference simulation (Zone III; Section~\ref{sec:zoneIII}). The streams most likely play a role in forming the rings and gaps observed in several other simulations as well, especially in the lower resistivity (D\_4) and stronger field (beta\_3) cases, where the disk evolution is dominated by a stream-induced envelope. The concept of avalanche accretion streams itself is not new; \citet{1998ApJ...508..186K} found such streams in their 2D simulations of thick AGN accretion disks (where the streams were termed `accretion avalanches'; see also \citealt{1994ApJ...433..746S,1996ApJ...461..115M,2009ApJ...707..428B}). What we have shown here is that such avalanche streams tend to develop in regions of low dimensionless magnetic diffusivity $D_m$ (or high Lundquist number $S=1/D_m$) and that they can deposit material at fixed locations quickly enough to form rings in some cases or generate enough spatial variation in mass accretion to produce multiple rings and gaps in others.

\subsection{Magnetic diffusion and wind-dominated vs. stream-dominated accretion}

We found two distinct modes of accretion in coupled disk-wind systems initially threaded by a relatively weak, ordered poloidal magnetic field. They are controlled by the dimensionless magnetic diffusivity parameter $D_m$ (or the Lundquist number $S=1/D_m$). Low values of $D_m$ promote the formation of avalanche accretion streams, leading to a `stream-dominated' mode of accretion. Higher values of $D_m$ suppress stream formation, leading to a more steady wind that drives accretion in a more laminar disk, i.e., a `wind-dominated' mode of accretion. If $D_m$ becomes too large, the strong wind dies out quickly as the wind-launching magnetic field weakens due to diffusion. 

The zero diffusivity limit has recently been explored in the 3D ideal MHD simulations of \citetalias{2017arXiv170104627Z} that came to our attention near the conclusion of this investigation. Their study has several aspects in common with the work presented here; for example, both concentrate on relatively thin disks (disk thickness $5-10$ per cent of the radius), both adopt a large-scale poloidal magnetic field initially corresponding to a typical plasma-$\beta$ of $10^3$ on the disk midplane, and both employ a spherical-polar coordinate system. The main differences are that (1) their simulations are fully 3D, whereas ours are 2D axisymmetric, and (2) their simulations are formally ideal MHD, whereas we included explicit resistivity. Despite these differences, there is qualitative agreement regarding some key features of the simulations. One of the most intriguing features of our 2D simulations is the development of fast accreting avalanche streams near the disk surface. \citetalias{2017arXiv170104627Z} also observes the simultaneous fast accretion at high altitudes and slow expansion on the midplane, where the fast accretion occurs in a thick layer above the disk extending up to $\sim 1-1.5$ times the local radius (which they term `disk corona'). We believe that their disk corona is physically equivalent to the thick envelope found in our low diffusivity ($D_{m,0}$) cases. In our case, the thick envelope is produced by the repeated formation and disruption of multiple avalanche accretion streams, as can be seen in Fig.~\ref{fig:edge_face} (also see movies in the online journal). Although it is not discussed explicitly in their work, we suspect that a similar mechanism is responsible for forming the corona in their simulations as well, although it may be harder to identify rapidly infalling streams in 3D simulations than in 2D ones. The avalanche streams tend to drag the field lines into a radial configuration, which are then wound up by differential rotation. This naturally produces the $r-\phi$ magnetic stress that is found by \citetalias{2017arXiv170104627Z} to dominate the coronal accretion. In any case, our simulations add weight to the emerging picture that a chaotic, fast accreting `corona' or envelope plays a crucial role in shaping the structure and evolution of thin disks threaded by relatively weak ($\beta \sim 10^3$), open magnetic field lines, especially when the magnetic diffusivity is low.

In more magnetically diffusive disks with larger values of $D_m$, the situation can be quite different. On one hand, the explicit diffusivity can enable outward field diffusion by itself, allowing for the possibility of a steady-state balance between the outward diffusion and the inward advection of field lines by mass accretion. On the other hand, it can suppress the MRI in the disk and thus remove the turbulent diffusivity that could also enable such a state. This field advection-diffusion balance is illustrated in Zone II of the reference run (Fig.~\ref{fig:path_0.3au}), where the disk connects smoothly to the wind. One might be tempted to identify the subsonic region at the base of the wind as the `corona', which has a similar height ($z\sim R$) in the simulations of \citetalias{2017arXiv170104627Z}. However, this subsonic wind base is expanding slowly, and the fast accretion (with an effective $\alpha$ parameter of $\sim 2$) is limited to the high density disk proper (see Fig.~\ref{fig:path_0.3au}(c)). Most of the accretion in the more diffusive cases is through the disk and is driven by the magnetic torque from the wind; this is in contrast to the ideal MHD simulations of \citetalias{2017arXiv170104627Z}, where the wind plays only a minor role in driving the accretion. Wind-driven accretion appears to hold for other 2D non-ideal MHD simulations of coupled disk-wind systems that include either resistivity (e.g., \citetalias{2016ApJ...825...14S}) or ambipolar diffusion and the Hall effect \citep{2017ApJ...836...46A}. Whether it stays true in 3D or not remains to be determined. In any case, our results, together with those of \citetalias{2017arXiv170104627Z} and others, suggest that explicit magnetic diffusion from non-ideal MHD effects plays a key role in determining the extent to which the accretion is dominated by a wind or by a thick `corona'/stream-induced envelope.

We speculate that the two modes of accretion may be unified in the following sense. For a thin disk magnetized with a relatively weak poloidal magnetic field, a toroidal magnetic field is quickly generated. This would naturally `puff up' the disk in the vertical direction, creating a thick envelope that is supported by the magnetic pressure gradient vertically (see also \citealt{2011ApJ...732L..30H}) and by rotation in the cylindrical $R$-direction. This rotationally supported envelope can become unstable to the MRI, just as the disk, especially when the dimensionless magnetic diffusivity $D_m$ is low. Indeed, its larger vertical extent makes the envelope more prone to the MRI than the thin disk itself. When the diffusivity $D_m$ is high, the MRI is suppressed in the envelope, which allows the magnetic pressure gradient-driven expansion that produced the envelope to continue smoothly to larger distances, forming a slow wind.

\subsection{Implications of ring and gap formation on dust dynamics and planet formation}

It is well-understood that an outward radial pressure gradient in a disk causes the gas to rotate at a sub-Keplerian velocity. Consequently, grains that orbit at the Keplerian speed would experience a headwind, lose angular momentum, and migrate inward  \citep{1977MNRAS.180...57W}. If, however, the radial pressure gradient is reversed due to a `bump' in the disk surface density, this local maximum is able to halt the inward radial drift of particles and trap them there. In 3D, these radial bumps in the disk surface density can lead to the growth of the Rossby wave instability (RWI), whereby a high pressure nonaxisymmetric vortex grows exponentially, creating an azimuthal dust trap \citep{1999ApJ...513..805L}. Azimuthally asymmetric features are observed in disks and can be explained by azimuthal variations in the gas-to-dust ratio and grain size segregation \citep{2013Natur.493..191C}. The formation of radial and azimuthal pressure traps are critical for stopping solid particles from spiraling into the star on rapid timescales, thereby allowing the onset of grain growth and the eventual formation of planetesimals. 

The emergence of multiple radial density maxima in the simulations presented here shows the potential importance that MHD disk winds and avalanche streams can have on growing the seeds of planet formation. This is especially true in the inner ($\lesssim1$~au) regions of the disks, where jets and winds of young stellar objects are most likely launched and where the most common type of planets (super-Earths/mini-Neptunes) are located. Indeed, Sun-like stars have a 50 per cent chance of having a compact system of small (non-giant) planets within 1~au \citep{2015ARA&A..53..409W}. While the importance of migration for this population of planets is still unknown, it seems plausible that many of these planets may have formed at their current locations. If this is the case, an efficient mechanism is needed for trapping dust grains on sub-au scales, especially during the relatively early phases of star formation when most of the mass is processed through the inner disk. Rings and gaps produced by variable magnetic winds and avalanche streams that drive rapid accretion potentially provide an opportunity for grains to accumulate and grow early in the disk's life. This mechanism will likely not be as efficient at later times when YSO jets and winds are observed to be less powerful and the accretion rate is lower. More work is needed to quantify the effects that the disk structures created by magnetically driven winds and avalanche streams have on the dynamics and growth of dust grains.

\subsection{Future directions}

The initial set of idealized simulations presented in this work is aimed at illustrating the basic processes through which rings and gaps are produced by variable winds and avalanche accretion streams. In future refinements we will extend the simulations to a less restrictive geometry. Preliminary 2D (axisymmetric) simulations with two hemispheres show qualitatively similar results to those presented here. Full 3D simulations, such as those of \citetalias{2017arXiv170104627Z} but with non-ideal MHD effects (see \citealt{2014ApJ...784..121S,2015ApJ...801...84G,2016arXiv161200883B}), will be needed to determine whether the formation of rings and gaps is artificially enhanced by the axisymmetric geometry and whether the rings formed in 2D simulations are unstable to the Rossby wave instability. Also, 3D simulations are needed to determine whether the accretion streams will survive in 3D since the streams are physically related to the MRI channel flows, which are known to behave differently in 2D and 3D \citep{1998RvMP...70....1B,2009MNRAS.394..715L,2015MNRAS.453.3257L}. Just as important, there is a need to include a detailed calculation of the ionization structure of the disk and the non-ideal MHD effects, especially the Hall effect, which is expected to be important near the disk midplane on the au scale \citep{2001ApJ...552..235B,2004MNRAS.348..355K}. Post-processing the simulations through radiative transfer calculations is needed to make predictions for observable quantities (e.g., variability in near infrared emission and profiles of optical forbidden lines) that can be compared directly with observations (e.g., \citealt{2014AJ....147...82C,2016ApJ...831..169S}). The long term goal is to include a self-consistent treatment of dust grains in the full 3D non-ideal MHD coupled disk-wind simulations. This would allow us to quantify the effects that wind and stream-induced disk structures have on grain trapping and growth, as well as any potential back-reaction that the grains might have on the gas dynamics of disk accretion, stream formation, and wind launching. The rings and gaps found in our starting simulations provide added impetus to explore the interplay between dust dynamics and winds and streams in greater depth.

\section{Conclusions}\label{sec:conc}

We have presented the results of 2D (axisymmetric) resistive MHD simulations of coupled disk-wind systems with a range of disk parameters (resistivity, magnetic field strength, and temperature), focusing on geometrically thin disks. We find that the structure and dynamics of the disk-wind system strongly depend on the dimensionless magnetic diffusivity parameter $D_m\equiv \eta/(h v_A)$ and that interesting disk features, including rings and gaps, are naturally produced. Specifically, we find that:

\begin{enumerate}
\item There are two distinct modes of accretion depending on the dimensionless parameter $D_m$. Disks with low values of $D_m$, from either a small resistivity or high field strength, tend to develop fast `avalanche accretion streams'. The rapid formation and disruption of such streams often leads to a clumpy, thick envelope above the disk that dominates the dynamics of the system (e.g., models beta\_3 and D\_4; see Fig.~\ref{fig:edge_face}), although a highly variable wind is still launched above the envelope. This envelope appears equivalent to the thick disk `corona' found independently by \citetalias{2017arXiv170104627Z} in their 3D ideal MHD simulations. In both cases, the disk below the corona/envelope is often expanding radially outward. The streams (and the thick clumpy envelope they produce) are suppressed in simulations with larger values of $D_m$ (from either a large resistivity or low field strength, e.g., models beta3 and D4). In these more diffusive (larger $D_m$) simulations, most of the accretion occurs through a laminar thin disk rather than the thick clumpy envelope, and the disk accretion is driven mainly by a magnetic wind.

\item Both wind-dominated and stream-dominated accretion create prominent features in the surface density distribution, especially rings and gaps. The wind-driven ring and gap formation is illustrated most clearly in the innermost region (Zone I) of the reference simulation, where there is substantial redistribution of magnetic flux relative to the mass in the disk that is enabled by the resistivity (Fig.~\ref{fig:mass2flux}). Regions with lower mass-to-flux ratios tend to drive stronger winds and accrete faster, producing gaps; those with higher mass-to-flux ratios tend to accrete more slowly, allowing matter to accumulate and form dense rings. Another clear illustration of wind-driven gap formation can be found in models D4 and beta3, where a strong wind is driven from the outer part of the disk but not from the more magnetically diffusive inner part, creating a deep gap between them (Fig.~\ref{fig:surfden_panels}(e) and (f)). The stream-driven ring formation is illustrated most clearly in the thicker disk model (t4), where a stream feeds a prominent ring at a roughly constant radius (Fig.~\ref{fig:edge_face}, right column). Rings formed this way also have high mass-to-flux ratios. Multiple rings and gaps are formed in other, more complicated cases, especially those with stream-induced envelopes (model D\_4 and beta\_3). It is likely that both magnetic winds and avalanche accretion streams play a role in their formation, although the relative importance of the two mechanisms is hard to quantify due to the complexity of the flow pattern inside and above the disk. 

\item Powerful winds are launched despite the fact that the magnetic field in the disk is rather weak initially (corresponding to a typical plasma-$\beta\sim10^3$). In the reference simulation where the wind is analyzed in detail, we find that the bulk of the wind is heavily mass-loaded and accelerated by the magnetic pressure gradient to relatively low speeds (a few~$\times10~\kms$). There are, however, lightly mass-loaded regions that are accelerated magnetocentrifugally to speeds exceeding $100~\kms$, comparable to the jet speeds observed in young stellar objects. The magnetic wind can remove angular momentum from the disk efficiently, leading to disk accretion with an effective $\alpha$ parameter up to order unity. Our simulations add weight to the notion of wind-driven disk evolution, especially in the presence of a suitable level of magnetic diffusivity.
\end{enumerate}

Rings and gaps produced in circumstellar disks by magnetic winds and avalanche accretion streams have important implications on the dynamics and growth of dust grains and ultimately planet formation. The local pressure maxima associated with the rings would act to stop the radial drift of solid particles, possibly trapping them long enough to enable enhanced grain growth that facilitates planetesimal formation. This may be especially important in the inner (i.e., few tenths of an au) disk regions where the largest population of planets reside, as seen by \textit{Kepler}. The simulations presented in this work lay the foundation for future explorations of these and other aspects of the coupled disk-wind systems.

\section*{Acknowledgements}

We thank Dom Pesce, Zhaohuan Zhu, and William B\'{e}thune for useful comments. This work is supported in part by NSF AST-1313083 and NASA NNX14AB38G. The simulations in this work were carried out on the Rivanna computer cluster at the University of Virginia.

\bibliography{main}{}
\bibliographystyle{mnras}

\bsp	
\label{lastpage}
\end{document}